\documentclass[11pt]{article}

\usepackage[a4paper,margin=1in]{geometry}
\usepackage[T1]{fontenc}
\usepackage[utf8]{inputenc}
\usepackage{lmodern}
\usepackage{amsmath,amssymb,mathtools,bm}
\usepackage{hyperref}
\usepackage{enumitem}
\usepackage{microtype}
\usepackage{graphicx}

\hypersetup{
  colorlinks=true,
  linkcolor=blue,
  citecolor=blue,
  urlcolor=blue,
  pdftitle={Constructive Cavity Method for Ising Restricted and Deep Boltzmann Machines with Gaussian Couplings},
  pdfauthor={Simone Franchini, Francesco Concetti, Giampiero Bardella, Liming Pan}
}

\newcommand{\E}{\mathbb{E}}
\newcommand{\cA}{\mathcal{A}}
\newcommand{\cP}{\mathcal{P}}

\newcommand{\Var}{\operatorname{Var}}

\title{Constructive Cavity Method for Ising Restricted and Deep Boltzmann Machines with Gaussian Couplings}

\author{%
\begin{tabular}{c}
Simone Franchini\textsuperscript{1,*}, Francesco Concetti\textsuperscript{2}\\
Giampiero Bardella\textsuperscript{3},
Liming Pan\textsuperscript{4}\\[0.6em]
\small \textsuperscript{1}~CNR-ISTC, Via Gian Domenico Romagnosi 18, 00196 Rome, Italy.\\
\small \textsuperscript{2}~Faculty of Mathematics and Computer Science, UniDistance Suisse, 3900 Brig, Switzerland.\\
\small \textsuperscript{3}~Department of Physiology and Pharmacology, Sapienza University of Rome,\\
\small Piazzale Aldo Moro 5, 00185 Roma, Italy.\\
\small \textsuperscript{4}~School of Cyber Science and Technology, University of Science and Technology of China,\\ 
\small Hefei 230026, China.\\[0.3em]
\small \textsuperscript{*}~Contact: \href{mailto:simone.franchini@yahoo.it}{simone.franchini@yahoo.it}
\end{tabular}%
}
\date{}

\usepackage{etoolbox}
\gappto\normalsize{%
  \setlength{\abovedisplayshortskip}{\abovedisplayskip}%
  \setlength{\belowdisplayshortskip}{\belowdisplayskip}%
}
\gappto\small{%
  \setlength{\abovedisplayshortskip}{\abovedisplayskip}%
  \setlength{\belowdisplayshortskip}{\belowdisplayskip}%
}
\gappto\footnotesize{%
  \setlength{\abovedisplayshortskip}{\abovedisplayskip}%
  \setlength{\belowdisplayshortskip}{\belowdisplayskip}%
}

\begin{document}

\maketitle

\begin{abstract}
We apply the Constructive Cavity Method (CCM) to the dense, Ising type, restricted and deep Boltzmann machine with Gaussian couplings between adjacent layers. After identifying the core and interface contributions, CCM identifies the blanket decomposition of the cavity variables with the Aizenman--Sims--Starr random overlap structure. We apply the Random Energy Model (REM) universality ansatz to represent the cavity free energies by those of a generalized random energy model (GREM) and evaluate them recursively through Poisson point processes (PPPs). In the layer-wise gauge used in this paper each layer is cut with its own cumulative fraction
\[
|U_\ell\cap L_p|=\lambda_p q_\ell^{(p)}N+o(N),
\qquad
\mathbf 0=\mathbf q_0\le \mathbf q_1\le\cdots\le \mathbf q_r=\mathbf 1,
\]
where \(\mathbf q_\ell\) is the vector valued parameter
\[
\mathbf q_\ell=(q_\ell^{(1)},\ldots,q_\ell^{(K)}).
\]
Here \(U_\ell\) is the cumulative blanket union up to blanket \(\ell\), and \(V_\ell:=U_\ell\setminus U_{\ell-1}\) is the \(\ell\)-th blanket increment. The index \(p\) labels architectural layers, whereas \(\ell\) labels the blankets. Thus, the construction is controlled by one layer-wise coordinate \(\mathbf q_\ell\) per blanket. The dependence on the multilayer geometry is carried by the two matrices
\[
M=2M_0\Lambda,
\qquad
M_1=\Lambda M_0\Lambda.
\]
The constructive cavity field in layer \(p\) has blanket variances
\[
Q_\ell^{(p)}-Q_{\ell-1}^{(p)}
=
\bigl(M(\mathbf q_\ell-\mathbf q_{\ell-1})\bigr)_p,
\]
while the correction field has blanket variances
\[
Q_\ell-Q_{\ell-1}
=
\mathbf q_\ell^\top M_1\mathbf q_\ell
-
\mathbf q_{\ell-1}^\top M_1\mathbf q_{\ell-1}.
\]
The nested REM--PPP average gives the finite-level constructive cavity functional
\[
\cP_{\mathrm{RBM/DBM}}^{\mathrm{CCM}}(\zeta,\mathbf q)
=
\log2+
\sum_{p=1}^K\lambda_pX_0^{(p)}
-
\frac12
\sum_{\ell=0}^{r-1}\zeta_\ell
\left(
\mathbf q_{\ell+1}^{\top}M_1\mathbf q_{\ell+1}
-
\mathbf q_\ell^{\top}M_1\mathbf q_\ell
\right).
\]
The variables \(\mathbf q_\ell\) are layer-wise cumulative fractions of the constructive gauge. The balanced scalar gauge is recovered by imposing \(\mathbf q_\ell=q_\ell\mathbf 1\). Here we derive the functional produced by the constructive cavity calculation; the selection of the physical value in the non-convex multilayer regime is a separate problem.
\end{abstract}

\section{Introduction}\label{sec:introduction}

The Constructive Cavity Method (CCM)~\cite{SimplifiedParisiIII} is a procedure for evaluating scalar cavity functionals by constructing the cavity variables explicitly. It originates from the Kernel/Blanket framework of disordered systems \cite{RSBWithoutReplicas,UniversalScaling,SimplifiedParisi} where replica-symmetry-breaking structures are represented through blanket decompositions, REM universality \cite{RSBWithoutReplicas,SimplifiedParisiII,REMUniversalityLinear,REMUniversalityLetter}, and constructive cavity variables. In the Sherrington--Kirkpatrick and Parisi spin-glass setting \cite{SherringtonKirkpatrick1975,Parisi1979,Guerra2003,Talagrand2006}, the method starts from the Aizenman--Sims--Starr cavity functional \cite{AizenmanSimsStarr2003}, identifies the blanket decomposition of its cavity variables with the random overlap structure (ROSt), and evaluates the averages recursively through Poisson point processes (PPPs).

The aim of this paper is to apply the same practical construction to the dense Ising restricted and deep Boltzmann machines with Gaussian couplings. The model has $K$ layers, interactions only between adjacent layers, and layer proportions
\[
|L_p|=\lambda_pN+o(N),
\qquad
\sum_{p=1}^K\lambda_p=1.
\]
The case $K=2$ is the dense bipartite Ising RBM with Gaussian couplings, while the case $K\ge3$ is the corresponding DBM architecture. Boltzmann machines and their restricted and deep variants originate as layered energy-based models \cite{AckleyHintonSejnowski1985,Smolensky1986,SalakhutdinovHinton2009}. Their statistical-mechanics and spin-glass formulations are connected with bipartite, multi-species, and multilayer mean-field models \cite{BarraGenoveseSollichTantari2017,BarraContucciMingioneTantari2015,Panchenko2015,AlbericiContucciMingione2021,AgliariAlbaneseAlemannoFachechi2021}. In non-convex multilayer regimes, the selection of the physical value is not expected to be governed automatically by a standard minimization principle \cite{ChenMourrat2025,Chen2026}. We do not address that selection problem here. We derive the constructive cavity functional.

The central constructive coordinate is a layer-wise gauge. For each blanket \(\ell\), each layer \(L_p\) is cut by its own cumulative fraction \(q_\ell^{(p)}\), weighted by the layer size:
\[
|U_\ell\cap L_p|=\lambda_p q_\ell^{(p)}N+o(N).
\]
Thus the finite-level construction uses one layer-wise coordinate
\[
\mathbf q_\ell=(q_\ell^{(1)},\ldots,q_\ell^{(K)})
\]
per blanket, with \(\mathbf q_\ell\) vector valued parameter and
\[
\mathbf 0=\mathbf q_0\le \mathbf q_1\le\cdots\le \mathbf q_r=\mathbf 1
\]
componentwise. The layer dependence is carried by the layer sizes, by the coupling strengths, and by the different components of \(\mathbf q_\ell\). In particular, if
\[
M_1=\Lambda M_0\Lambda,
\qquad
M=2M_0\Lambda,
\]
where \(M_0\) is the weighted adjacency matrix of the layer chain and \(\Lambda=\operatorname{diag}(\lambda_1,\ldots,\lambda_K)\), then the cavity field acting on a spin in layer \(p\) has blanket variances
\[
\bigl(M(\mathbf q_\ell-\mathbf q_{\ell-1})\bigr)_p,
\]
while the correction field has blanket variances
\[
\mathbf q_\ell^\top M_1\mathbf q_\ell
-
\mathbf q_{\ell-1}^\top M_1\mathbf q_{\ell-1}.
\]
The balanced scalar gauge is obtained as the diagonal specialization
\[
\mathbf q_\ell=q_\ell\mathbf 1.
\]
In that case the general matrix quantities reduce to
\[
d_p=(M\mathbf 1)_p
=2\bigl(\beta_{p-1}^2\lambda_{p-1}+\beta_p^2\lambda_{p+1}\bigr),
\]
\[
\kappa=\mathbf 1^\top M_1\mathbf 1
=2\sum_{p=1}^{K-1}\beta_p^2\lambda_p\lambda_{p+1}.
\]

We first identify cores and interfaces in the layer-wise partition and compute the Gaussian blanket amplitudes. The REM universality ansatz of Ref.~\cite{SimplifiedParisiIII}, applied in Section~\ref{sec:ass-substitution}, represents jointly the \(K\) cavity free-energy shifts and the correction shift by their generalized random energy model (GREM) counterparts. The blanket components are identified with the Aizenman--Sims--Starr ROSt variables, and nested Poisson averaging yields the finite-level functional.
The final finite-level functional is
\begin{equation}\label{eq:intro-functional-scalar}
\cP_{\mathrm{RBM/DBM}}^{\mathrm{CCM}}(\zeta,\mathbf q)
=
\log2+
\sum_{p=1}^K\lambda_pX_0^{(p)}
-
\frac12
\sum_{\ell=0}^{r-1}
\zeta_\ell
\left(
\mathbf q_{\ell+1}^{\top}M_1\mathbf q_{\ell+1}
-
\mathbf q_\ell^{\top}M_1\mathbf q_\ell
\right).
\end{equation}
The quantities \(X_0^{(p)}\) are obtained by the nested REM--PPP Gaussian recursion with layer-dependent scale \((M(\mathbf q_\ell-\mathbf q_{\ell-1}))_p\) at blanket level \(\ell\). The layer-wise cumulative fractions \(\mathbf q_\ell\) determine the amplitudes and supply the levels of the resulting Parisi-type functional.

The paper is organized as follows. Section~\ref{sec:ccm-practice} recalls the CCM in the operational form used here. Section~\ref{sec:model} defines the Ising RBM/DBM with Gaussian couplings and the matrices \(M\) and \(M_1\). Section~\ref{sec:io-balanced} separates cores and interfaces and introduces the layer-wise gauge. The following sections construct the Gaussian variables, write the Aizenman--Sims--Starr functional, perform the nested REM--PPP average, and record basic checks. Section~\ref{sec:full-rsb-comparisons} relates the result to the bipartite and multi-species full-RSB formulations, and Section~\ref{sec:why-ccm-works} explains the construction.

\section{The CCM in practice}\label{sec:ccm-practice}

We use the minimal construction of Ref.~\cite[Section~4.7]{SimplifiedParisiIII}; Section~\ref{sec:why-ccm-works} discusses its theoretical background.

The CCM~\cite{SimplifiedParisiIII} computes cavity free energies by decomposing the cavity variables into blanket components and evaluating their normalized averages. In a single-population setting the scalar cavity functional has the schematic form
\[
\cA(\nu)
=
\log2+
\E\log\langle C(x)\rangle_\nu
-
\E\log\langle F(y)\rangle_\nu,
\]
where $x$ is the cavity field, $y$ is the correction or fugacity field, and $\nu$ is a reservoir. This is the Aizenman--Sims--Starr type of scalar cavity functional \cite{AizenmanSimsStarr2003}, evaluated in the constructive form of the CCM \cite{SimplifiedParisiIII}. For Ising spins,
\[
C(x)=\cosh(h+x),
\qquad
F(y)=e^y.
\]

For a multilayer system there is one cavity mark for each layer. Any spin added to layer $p$ has a cavity field denoted by $x_p$. The multilayer scalar functional will therefore have the form
\[
\log2+
\sum_{p=1}^K\lambda_p
\E\log\langle \cosh(h^{(p)}+x_p)\rangle_\nu
-
\E\log\langle e^y\rangle_\nu.
\]
The coefficient $\lambda_p$ appears because the pressure is normalized per total spin, while the layer $p$ contains a fraction $\lambda_p$ of all spins.
In its minimal form, CCM starts from the Aizenman--Sims--Starr cavity
functional, applies the blanket decomposition to its cavity and
correction variables, and evaluates the resulting averages by nested
Poisson averaging. Equivalently, CCM identifies the blanket decomposition
of the cavity variables with the Aizenman--Sims--Starr random overlap
structure (ROSt). Ref.~\cite[Sections~3 and~4]{SimplifiedParisiIII} provides a physical justification of this identification through the GREM representation of the states and the joint construction of their masses and cavity fields.

For RBM/DBM, we first identify the cores and interfaces of the layer-wise
partition. The adjacent-layer vertex counts determine the cavity
amplitudes, and the complete edge-shell counts determine the correction
amplitudes. We then use these amplitudes in the
thermodynamic representation stated in Section~\ref{sec:ass-substitution}
and perform the nested average of Section~\ref{sec:ppp}.

\section{The Ising RBM/DBM model with Gaussian couplings}\label{sec:model}

Fix a finite number $K\ge2$ of layers $L_1,\ldots,L_K$. Let $N_p=|L_p|$ and
\[
N=\sum_{p=1}^K N_p,
\qquad
\frac{N_p}{N}\to \lambda_p>0,
\qquad
\sum_{p=1}^K\lambda_p=1.
\]
At each site $i\in L_p$ there is an Ising spin
\[
\sigma_i^{(p)}\in\{-1,+1\}.
\]
The Hamiltonian of the RBM/DBM is
\begin{equation}\label{eq:hamiltonian}
H_N(\sigma)
=
\frac{\sqrt2}{\sqrt N}
\sum_{p=1}^{K-1}\beta_p
\sum_{i\in L_p}\sum_{j\in L_{p+1}}
J_{ij}^{(p)}\sigma_i^{(p)}\sigma_j^{(p+1)},
\end{equation}
where the variables $J_{ij}^{(p)}$ are independent standard Gaussians and $\beta_p\ge0$ is the coupling strength between layers $p$ and $p+1$. The case $K=2$ is a dense bipartite Ising RBM with Gaussian couplings. The case $K\ge3$ is the corresponding DBM with interactions only between adjacent layers.
Introduce the diagonal matrix
\[
\Lambda=\operatorname{diag}(\lambda_1,\ldots,\lambda_K).
\]
We write \(\mathbf1\) for the \(K\)-dimensional vector of ones.
Let $M_0$ be the symmetric tridiagonal matrix defined by the following formula:
\[
(M_0)_{p,p+1}=(M_0)_{p+1,p}=\beta_p^2,
\qquad p=1,\ldots,K-1,
\]
with all other entries equal to zero. We use
\[
M_1=\Lambda M_0\Lambda,
\qquad
M=2M_0\Lambda.
\]
Equivalently, for a vector \(v=(v^{(1)},\ldots,v^{(K)})\),
\[
(Mv)_p
=
2\beta_{p-1}^2\lambda_{p-1}v^{(p-1)}
+
2\beta_p^2\lambda_{p+1}v^{(p+1)},
\]
where the terms with indices outside \(1,\ldots,K\) are omitted. This convention also applies to the boundary layers in the formulas below.
In the balanced specialization used for comparison later, the two scalar constants are
\begin{equation}\label{eq:dp-def}
d_p=(M\mathbf1)_p
=
2\bigl(\beta_{p-1}^2\lambda_{p-1}+\beta_p^2\lambda_{p+1}\bigr),
\end{equation}
\begin{equation}\label{eq:kappa-def}
\kappa=\mathbf1^\top M_1\mathbf1
=
2\sum_{p=1}^{K-1}\beta_p^2\lambda_p\lambda_{p+1}.
\end{equation}
The factor $\sqrt2$ in \eqref{eq:hamiltonian} is chosen so that these expressions can have this simple symmetric form. External fields are not written in the Hamiltonian. Throughout the derivation, $h^{(p)}$ denotes an optional deterministic field for layer $p$ entering only the one-spin cavity mark. Setting $h^{(p)}=0$ recovers exactly the Hamiltonian in \eqref{eq:hamiltonian}.

\section{Interface-only structure and layer-wise gauge}\label{sec:io-balanced}

We first separate the core and interface contributions of the layer-wise partition, and use interface dominance under refinement to obtain the interface-only (IO) representative. Here and below, \(p\) labels the architectural layers \(L_p\), while \(\ell\) labels the blankets of the constructive partition; \(\zeta_{\ell-1}\) is the Poisson exponent used to average blanket \(\ell\), and \(q_\ell^{(p)}\) is the fraction of layer \(L_p\) included up to blanket \(\ell\). We fix ordered disjoint blankets \(V_1,\ldots,V_r\), independently of the couplings, and set
\[
U_0=\varnothing,
\qquad
U_\ell:=\bigcup_{s=1}^{\ell}V_s,
\qquad
U_r=\bigcup_{p=1}^K L_p.
\]
Thus \(U_\ell\) is the cumulative blanket up to \(\ell\), and \(V_\ell=U_\ell\setminus U_{\ell-1}\) is the \(\ell\)-th blanket increment. This is the ordered blanket decomposition used in the CCM~\cite{SimplifiedParisiIII}. Each blanket \(V_\ell\) is a global CCM block crossing all architectural layers, with
\[
V_\ell=\bigsqcup_{p=1}^K(V_\ell\cap L_p).
\]
We choose a layer-wise nondecreasing profile
\[
\mathbf 0=\mathbf q_0\le \mathbf q_1\le\cdots\le \mathbf q_r=\mathbf 1,
\qquad
\mathbf q_\ell=(q_\ell^{(1)},\ldots,q_\ell^{(K)}),
\]
where the inequalities are componentwise. The layer-wise constructive gauge cuts each architectural layer \(L_p\) by the cumulative fraction \(q_\ell^{(p)}=|U_\ell\cap L_p|/|L_p|+o(1)\), weighted by the layer size: therefore the cumulative blocks have size
\begin{equation}\label{eq:balanced-cumulative}
|U_\ell\cap L_p|
=
\lambda_p q_\ell^{(p)}N+o(N),
\qquad p=1,\ldots,K.
\end{equation}
The corresponding blanket block is
\[
V_\ell\cap L_p=(U_\ell\cap L_p)\setminus(U_{\ell-1}\cap L_p),
\]
and therefore its cardinality is
\begin{equation}\label{eq:balanced-block}
|V_\ell\cap L_p|
=
\lambda_p\bigl(q_\ell^{(p)}-q_{\ell-1}^{(p)}\bigr)N+o(N).
\end{equation}

\begin{figure}[t]
\centering
\includegraphics[width=\textwidth]{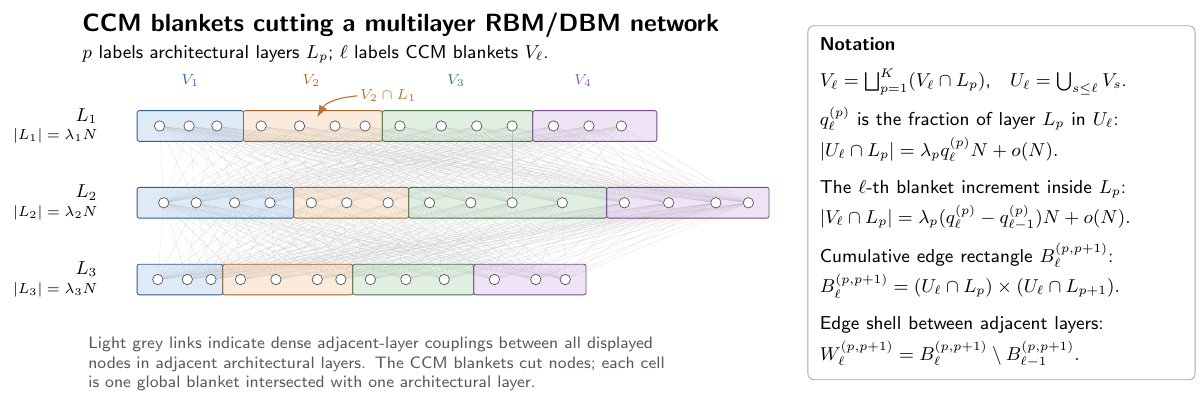}
\caption{CCM blanket geometry for a multilayer RBM/DBM. Architectural layers are indexed by \(p\), while CCM blankets are indexed by \(\ell\). Each blanket \(V_\ell\) is a global partition block crossing all architectural layers, and the cell \(V_\ell\cap L_p\) has size \(\lambda_p(q_\ell^{(p)}-q_{\ell-1}^{(p)})N+o(N)\).}
\label{fig:ccm-blanket-network-geometry}
\end{figure}

The variables \(\mathbf q_\ell\) are layer-wise cumulative fractions of the constructive gauge. The balanced scalar gauge is recovered by imposing \(\mathbf q_\ell=q_\ell\mathbf 1\).

The adjacent-layer structure of \eqref{eq:hamiltonian} determines the cavity components as follows. A cavity spin inserted in layer \(p\) interacts only with spins in the neighboring layers \(L_{p-1}\) and \(L_{p+1}\). Therefore the variance of its blanket-\(\ell\) cavity field is determined by the vertex counts in
\[
V_\ell\cap L_{p-1},
\qquad
V_\ell\cap L_{p+1}.
\]

The correction field has the same adjacent-layer interaction support and is decomposed into complete edge shells~\cite{RSBWithoutReplicas,UniversalScaling}. For adjacent layers \(L_p\) and \(L_{p+1}\), define the cumulative rectangle
\[
B_\ell^{(p,p+1)}=(U_\ell\cap L_p)\times(U_\ell\cap L_{p+1}).
\]
The blanket-\(\ell\) edge shell is
\[
W_\ell^{(p,p+1)}
=
B_\ell^{(p,p+1)}\setminus B_{\ell-1}^{(p,p+1)},
\qquad \ell=1,\ldots,r.
\]
Using \eqref{eq:balanced-cumulative}, its cardinality is
\begin{equation}\label{eq:balanced-shell}
|W_\ell^{(p,p+1)}|
=
\lambda_p\lambda_{p+1}N^2
\left(
q_\ell^{(p)}q_\ell^{(p+1)}
-
q_{\ell-1}^{(p)}q_{\ell-1}^{(p+1)}
\right)
+o(N^2).
\end{equation}
At finite level, the shell \(B_\ell^{(p,p+1)}\setminus B_{\ell-1}^{(p,p+1)}\) decomposes into an interface part and a core part. This decomposition introduces no additional datum. If
\[
\Delta q_\ell^{(p)}
=
q_\ell^{(p)}-q_{\ell-1}^{(p)},
\]
then we have
\[
q_\ell^{(p)}q_\ell^{(p+1)}
-
q_{\ell-1}^{(p)}q_{\ell-1}^{(p+1)}
=
q_{\ell-1}^{(p)}\Delta q_\ell^{(p+1)}
+
q_{\ell-1}^{(p+1)}\Delta q_\ell^{(p)}
+
\Delta q_\ell^{(p)}\Delta q_\ell^{(p+1)}.
\]
The first two terms are the interface contribution, while the last is the core contribution. Refining the layer-wise partition makes the total core fraction vanish. The IO pressure reduction follows from Eq.~\eqref{eq:dbm-core-pressure-bound}. The interface-dominance argument and the order of limits are discussed in Ref.~\cite[Section~2.1]{SimplifiedParisiIII}. We retain the complete finite edge-shell counts when constructing the correction field below.

The counts \eqref{eq:balanced-block} and \eqref{eq:balanced-shell} determine the Gaussian amplitudes of the constructive cavity variables.

\section{Constructive cavity variables}\label{sec:constructive-variables}

The blanket decomposition of Section~\ref{sec:io-balanced} supplies the two sets of counts required by CCM: vertices in neighboring-layer blocks for the cavity fields, and edges in adjacent-layer shells for the correction field. The Gaussian amplitudes depend only on the layer-wise profile \(\mathbf q\), the layer proportions \(\lambda\), and the coupling strengths \(\beta\). We use these amplitudes to realize the cavity variables on the REM--GREM reservoir of Sections~\ref{sec:ass-substitution}--\ref{sec:ppp}.

\subsection*{The cavity field}

Consider a new spin inserted in layer \(p\). It interacts only with the two neighboring layers \(L_{p-1}\) and \(L_{p+1}\). The random part of its cavity field is
\begin{equation}\label{eq:xNp-scalar}
x_{N,p}(\sigma)
=
\frac{\sqrt2}{\sqrt N}
\left(
\beta_{p-1}
\sum_{i\in L_{p-1}}
\widetilde J_i^{(p-1)}\sigma_i^{(p-1)}
+
\beta_p
\sum_{j\in L_{p+1}}
\widehat J_j^{(p)}\sigma_j^{(p+1)}
\right),
\end{equation}
with the missing boundary term omitted. The Gaussian variables in \eqref{eq:xNp-scalar} are fresh and independent of the reservoir.

The blanket-\(\ell\) contribution is obtained by restricting the sums to the blanket blocks in the neighboring layers:
\begin{equation}\label{eq:xNp-level-scalar}
x_{N,p,\ell}(\sigma)
=
\frac{\sqrt2}{\sqrt N}
\left(
\beta_{p-1}
\sum_{i\in V_\ell\cap L_{p-1}}
\widetilde J_i^{(p-1)}\sigma_i^{(p-1)}
+
\beta_p
\sum_{j\in V_\ell\cap L_{p+1}}
\widehat J_j^{(p)}\sigma_j^{(p+1)}
\right).
\end{equation}
Conditionally on the reservoir configuration, the cross terms vanish because the fresh couplings are independent and centered. Since the spins are Ising variables, the conditional variance is exactly computed as follows:
\begin{equation}\label{eq:x-level-raw-var-scalar}
\Var(x_{N,p,\ell}\mid\sigma)
=
\frac{2\beta_{p-1}^2}{N}|V_\ell\cap L_{p-1}|
+
\frac{2\beta_p^2}{N}|V_\ell\cap L_{p+1}|.
\end{equation}
At the boundary layers, the absent term is omitted.
Using the layer-wise block sizes \eqref{eq:balanced-block}, this becomes
\begin{align}\label{eq:x-level-var-scalar}
\Var(x_{N,p,\ell}\mid\sigma)
&=
2\beta_{p-1}^2\lambda_{p-1}
\bigl(q_\ell^{(p-1)}-q_{\ell-1}^{(p-1)}\bigr)
+
2\beta_p^2\lambda_{p+1}
\bigl(q_\ell^{(p+1)}-q_{\ell-1}^{(p+1)}\bigr)
+o(1)
\notag\\
&=
\bigl(M(\mathbf q_\ell-\mathbf q_{\ell-1})\bigr)_p+o(1).
\end{align}
Write \(\alpha=(\alpha_1,\ldots,\alpha_r)\) for a path of the REM--GREM representation. Its depth-\(\ell\) prefix is \(\alpha_{\leq\ell}=(\alpha_1,\ldots,\alpha_\ell)\). With the amplitude in \eqref{eq:x-level-var-scalar}, the constructive cavity field in layer \(p\) is
\begin{equation}\label{eq:xp-constructive-scalar}
x_p^\alpha
=
\sum_{\ell=1}^{r}
z_{\ell,p}(\alpha_{\leq\ell})
\sqrt{
\bigl(M(\mathbf q_\ell-\mathbf q_{\ell-1})\bigr)_p
},
\end{equation}
where distinct nodes, levels and architectural layers carry independent standard Gaussians, independent of the reservoir weights. Paths through the same node share its Gaussian variable, as in Ref.~\cite[Section~4.5]{SimplifiedParisiIII}. Equivalently, if
\begin{equation}\label{eq:Qp-scalar}
Q_\ell^{(p)}=(M\mathbf q_\ell)_p,
\end{equation}
\[
x_p^\alpha
=
\sum_{\ell=1}^{r}
z_{\ell,p}(\alpha_{\leq\ell})
\sqrt{Q_\ell^{(p)}-Q_{\ell-1}^{(p)}}.
\]
Thus the layer-wise profile \(\mathbf q\) and the matrix \(M\) fix the amplitudes of the layer-\(p\) cavity field.

\subsection*{The correction field}

The correction, or fugacity, mark accounts for the internal adjacent-layer interactions compensated in the cavity normalization. For adjacent layers \(L_p\) and \(L_{p+1}\), the blanket-\(\ell\) edge shell is \(W_\ell^{(p,p+1)}\) from \eqref{eq:balanced-shell}. The corresponding Gaussian correction mark is
\begin{equation}\label{eq:yN-level-scalar}
y_{N,\ell}^{(p,p+1)}(\sigma)
=
\frac{\sqrt2\,\beta_p}{N}
\sum_{(i,j)\in W_\ell^{(p,p+1)}}
\overline J_{ij}^{(p)}
\sigma_i^{(p)}\sigma_j^{(p+1)}.
\end{equation}
The normalization is \(1/N\), rather than \(1/\sqrt N\), because the correction mark is an internal edge sum over \(O(N^2)\) terms and must have variance of order one.

Again, conditionally on the reservoir, the cross terms vanish by independence and centering of the fresh Gaussians, while each diagonal term contributes one because the spins are Ising. Hence, the expression for the variance is
\begin{equation}\label{eq:y-level-raw-var-scalar}
\Var\bigl(y_{N,\ell}^{(p,p+1)}\mid\sigma\bigr)
=
\frac{2\beta_p^2}{N^2}|W_\ell^{(p,p+1)}|.
\end{equation}
Using \eqref{eq:balanced-shell}, we get
\begin{equation}\label{eq:y-level-var-p-scalar}
\Var\bigl(y_{N,\ell}^{(p,p+1)}\mid\sigma\bigr)
=
2\beta_p^2\lambda_p\lambda_{p+1}
\left(
q_\ell^{(p)}q_\ell^{(p+1)}
-
q_{\ell-1}^{(p)}q_{\ell-1}^{(p+1)}
\right)
+o(1).
\end{equation}
The edge-shell contributions use independent fresh couplings, so the blanket variances add over \(p\). Summing over the adjacent layer pairs gives
\begin{equation}\label{eq:y-level-var-scalar}
Q_\ell-Q_{\ell-1}
=
\mathbf q_\ell^\top M_1\mathbf q_\ell
-
\mathbf q_{\ell-1}^\top M_1\mathbf q_{\ell-1}.
\end{equation}
Equivalently, we can write it in a compact matrix notation
\begin{equation}\label{eq:Q-scalar}
Q_\ell=\mathbf q_\ell^\top M_1\mathbf q_\ell.
\end{equation}
Since the profile is componentwise nondecreasing and \(M_1\) has nonnegative off-diagonal entries along the layer chain, the increments \(Q_\ell-Q_{\ell-1}\) are nonnegative.

The constructive correction field is therefore
\begin{equation}\label{eq:y-constructive-scalar}
y^\alpha
=
\sum_{\ell=1}^{r}
g_\ell(\alpha_{\leq\ell})
\sqrt{
\mathbf q_\ell^\top M_1\mathbf q_\ell
-
\mathbf q_{\ell-1}^\top M_1\mathbf q_{\ell-1}
},
\end{equation}
with standard Gaussian variables \(g_\ell(\alpha_{\leq\ell})\), independent between distinct nodes and of all cavity-field Gaussians and reservoir weights. Paths through the same prefix share the same correction variable.

\section{Aizenman--Sims--Starr cavity functional and constructive substitution}\label{sec:ass-substitution}

We now write the cavity functional using the blanket components constructed above. Let \(\nu_N\) be a trial reservoir on configurations of the existing \(K\)-layer system. For the Gibbs reservoir, the Aizenman--Sims--Starr construction compares the partition functions before and after adding one cavity spin; the same expression defines the functional on trial reservoirs. If the added spin belongs to architectural layer \(L_p\), summing its Ising value gives
\[
\sum_{\varepsilon=\pm1}
\exp\left\{\varepsilon\bigl(h^{(p)}+x_{N,p}\bigr)\right\}
=
2\cosh\bigl(h^{(p)}+x_{N,p}\bigr).
\]
Since the architectural layer \(L_p\) contributes a fraction \(\lambda_p\) of all spins, the cavity terms are averaged with weights \(\lambda_p\), and the factors \(2\) give the total entropy contribution \(\sum_p\lambda_p\log2=\log2\). The fugacity mark \(y_N\) compensates the internal adjacent-layer interactions of the reservoir and therefore appears with a minus sign. Thus the Aizenman--Sims--Starr-type scalar cavity functional associated with the multilayer model has the form \cite{AizenmanSimsStarr2003,SimplifiedParisiIII}
\begin{equation}\label{eq:ass-functional-scalar}
\cA_N(\nu_N)
=
\log2+
\sum_{p=1}^K\lambda_p
\E\log
\left\langle
\cosh\bigl(h^{(p)}+x_{N,p}\bigr)
\right\rangle_{\nu_N}
-
\E\log
\left\langle e^{y_N}\right\rangle_{\nu_N}.
\end{equation}
Here \(x_{N,p}\) is the one-spin cavity field associated with adding a spin to layer \(p\), and \(y_N\) is the Aizenman--Sims--Starr fugacity or correction mark. The coefficient \(\lambda_p\) appears because the pressure is normalized per total spin and layer \(p\) contains a fraction \(\lambda_p\) of all spins. The single entropy term is \(\log2\) because
\[
\sum_{p=1}^K \lambda_p\log2=\log2.
\]
The outer expectation is over the fresh Gaussian variables used in the cavity marks and over any randomness of the reservoir. The brackets denote averaging with respect to the reservoir.

For the physical calculation, we apply the REM universality ansatz of
Ref.~\cite[Sections~3 and~4.5]{SimplifiedParisiIII}: the \(K\) logarithmic
cosh averages and the logarithmic correction average in
\eqref{eq:ass-functional-scalar} are represented asymptotically in joint
distribution by the corresponding GREM free-energy shifts, with the
cavity variables below. On a physical fibre, which fixes the
magnetizations of the nonempty cells \(V_\ell\cap L_p\), the blanket
average uses complete Gibbs conditional kernels, defined as ratios of full
marginals. Contributions of retained fibres are summed in each partition
function before taking its logarithm~\cite[Section~2.4]{SimplifiedParisiIII}. The disorder expectation follows these operations.
This thermodynamic replacement uses the constructive fields
\[
x_{N,p}
\quad\rightsquigarrow\quad
x_p^\alpha
=
\sum_{\ell=1}^{r}
z_{\ell,p}(\alpha_{\leq\ell})
\sqrt{
\bigl(M(\mathbf q_\ell-\mathbf q_{\ell-1})\bigr)_p
},
\]
\[
y_N
\quad\rightsquigarrow\quad
y^\alpha
=
\sum_{\ell=1}^{r}
g_\ell(\alpha_{\leq\ell})
\sqrt{
\mathbf q_\ell^\top M_1\mathbf q_\ell
-
\mathbf q_{\ell-1}^\top M_1\mathbf q_{\ell-1}
}.
\]
Here \(\rightsquigarrow\) denotes this replacement in the cavity free-energy calculation. The realization by prefix-indexed variables follows Ref.~\cite[Section~4.5]{SimplifiedParisiIII}.

Thus, for a generic constructive reservoir \(\nu\), the finite-level constructive functional associated with the layer-wise profile \(\mathbf q\) is
\begin{equation}\label{eq:constructive-functional-scalar}
\cA_{r,\mathbf q}^{\mathrm{constr}}(\nu)
=
\log2+
\sum_{p=1}^K\lambda_p
\E\log
\left\langle
\cosh\bigl(h^{(p)}+x_p^\alpha\bigr)
\right\rangle_{\nu}
-
\E\log
\left\langle e^{y^\alpha}\right\rangle_{\nu}.
\end{equation}
The matrices \(M,M_1\) and the layer-wise profile \(\mathbf q\) determine the amplitudes in this functional. We now evaluate it on the REM--GREM reservoir.

\section{REM--PPP evaluation}\label{sec:ppp}

We evaluate \eqref{eq:constructive-functional-scalar} on the hierarchical
REM--PPP reservoir of Ref.~\cite[Sections~3.5--3.6]{SimplifiedParisiIII},
with parameters
\[
0=\zeta_{-1}\le \zeta_0\le\cdots\le\zeta_{r-1}\le \zeta_r=1.
\]
First take \(0<\zeta_0<\cdots<\zeta_{r-1}<1\). At every parent
\(u=(\alpha_1,\ldots,\alpha_{\ell-1})\), draw a PPP
\(\{w_\ell(u,i)\}_{i\ge1}\) with intensity
\(\zeta_{\ell-1}t^{-1-\zeta_{\ell-1}}\,dt\) on \((0,\infty)\).
Different parents and levels carry independent PPPs. The normalized
path weights and their average are
\[
\xi_\alpha=
\frac{\prod_{\ell=1}^{r}w_\ell(\alpha_{\leq\ell})}
{\sum_\gamma\prod_{\ell=1}^{r}w_\ell(\gamma_{\leq\ell})},
\qquad
\langle f^\alpha\rangle_\xi=\sum_\alpha\xi_\alpha f^\alpha.
\]
The increasing exponents make the total path mass finite almost surely.
Each parent mass includes all its descendants. These are the Ruelle
cascade weights~\cite{Ruelle1987,BolthausenSznitman1998,Arguin2006}, with the
Gaussian innovations of Section~\ref{sec:constructive-variables}
independent of the weights.

The marked Poisson identity acts at each parent on positive insertions
and their complete descendant sums. For exponent \(m\in(0,1)\), masses
\(w_i\), and i.i.d.\ positive pairs \((U_i,V_i)\) independent of the
masses, it gives
\[
\E\log\frac{\sum_iw_iU_i}{\sum_iw_iV_i}
=\frac{\log\E U^m-\log\E V^m}{m},
\qquad \E U^m,\E V^m<\infty.
\]
The rescaled totals have the same marginal law, so their expected
logarithms cancel. The recursion with descendant totals is given in
Ref.~\cite[Section~4.6]{SimplifiedParisiIII}.
In the multilayer functional the common normalization has coefficient
\(\sum_p\lambda_p-1=0\). The expectation is over the reservoir and its
Gaussian variables, after the state sums and logarithms.

Because the cavity part of \eqref{eq:constructive-functional-scalar} is a sum of layer-wise log-brackets, the Gaussian recursion is applied separately to each layer \(p\), with the same hierarchy \(\zeta\). Define the terminal value
\begin{equation}\label{eq:Xr-scalar}
X_r^{(p)}
=
\log\cosh\left(
h^{(p)}
+
\sum_{k=1}^{r}
\eta_{k,p}
\sqrt{
\bigl(M(\mathbf q_k-\mathbf q_{k-1})\bigr)_p
}
\right),
\end{equation}
where the \(\eta_{k,p}\) are independent standard Gaussians. For \(\ell=r-1,\ldots,0\), set
\begin{equation}\label{eq:X-recursion-scalar}
X_\ell^{(p)}
=
\frac1{\zeta_\ell}
\log
\E_{\ell+1}
\exp\left(\zeta_\ell X_{\ell+1}^{(p)}\right),
\end{equation}
where \(\E_{\ell+1}\) integrates the Gaussian increment at blanket \(\ell+1\) for the fixed layer \(p\), with exponent \(\zeta_\ell\). The formulas extend continuously to the boundary parameters. If \(\zeta_\ell=0\), the limiting step is
\[
X_\ell^{(p)}=\E_{\ell+1}X_{\ell+1}^{(p)}.
\]
Any initial zero-exponent blocks have fixed prefix labels, so their Gaussian increments in both fields are shared by all states. Together they represent the common overlap component.
For \(\zeta_\ell=1\), the step is
\(X_\ell^{(p)}=\log\E_{\ell+1}e^{X_{\ell+1}^{(p)}}\), the thermal
average. Consecutive equal exponents merge by Gaussian convolution;
\(\zeta_r=1\) records the terminal convention. The nested REM--PPP
average of the layer-\(p\) cavity mark is then
\begin{equation}\label{eq:cavity-ppp-value-scalar}
\E\log
\left\langle
\cosh\bigl(h^{(p)}+x_p^\alpha\bigr)
\right\rangle_\xi
=
X_0^{(p)}.
\end{equation}

The correction mark is evaluated by the same stability identity in its linear Gaussian form. Since \(y^\alpha\) is centered and has blanket-\(k\) variance increment
\[
\mathbf q_k^\top M_1\mathbf q_k
-
\mathbf q_{k-1}^\top M_1\mathbf q_{k-1},
\]
we get the fugacity contribution
\begin{equation}\label{eq:correction-ppp-value-scalar}
\E\log
\left\langle e^{y^\alpha}\right\rangle_\xi
=
\frac12
\sum_{\ell=0}^{r-1}
\zeta_\ell
\left(
\mathbf q_{\ell+1}^\top M_1\mathbf q_{\ell+1}
-
\mathbf q_\ell^\top M_1\mathbf q_\ell
\right).
\end{equation}
Indeed, at one PPP level the identity is
\[
\frac1m\log \E\exp\left(m g\sqrt{v}\right)
=
\frac{m}{2}v,
\]
with the value \(0\) when \(m=0\) by continuity.
Subdivision at a fixed averaging exponent preserves this correction, since the differences of the quadratic form telescope.
Substituting \eqref{eq:cavity-ppp-value-scalar} and \eqref{eq:correction-ppp-value-scalar} into \eqref{eq:constructive-functional-scalar} gives the finite-level CCM functional
\begin{equation}\label{eq:final-functional-scalar}
\cP_{\mathrm{RBM/DBM}}^{\mathrm{CCM}}(\zeta,\mathbf q)
=
\log2+
\sum_{p=1}^K\lambda_p X_0^{(p)}
-
\frac12
\sum_{\ell=0}^{r-1}
\zeta_\ell
\left(
\mathbf q_{\ell+1}^{\top}M_1\mathbf q_{\ell+1}
-
\mathbf q_\ell^{\top}M_1\mathbf q_\ell
\right).
\end{equation}
This is the layer-wise constructive cavity functional of the multilayer RBM/DBM model. The variables \(\mathbf q_\ell\) entered the derivation as layer-wise cumulative fractions of the constructive gauge. The balanced scalar gauge is recovered by imposing \(\mathbf q_\ell=q_\ell\mathbf 1\).
An equivalent representation by independent PPPs, with parent indices
carrying the shared cavity variables, gives the same random cavity
functional in distribution~\cite[Sections~3.6 and~4.6]{SimplifiedParisiIII}.

\section{Checks}\label{sec:checks}

We present consistency checks of the layer-wise functional \eqref{eq:final-functional-scalar}. The balanced scalar formulas are recovered by imposing \(\mathbf q_\ell=q_\ell\mathbf 1\). These checks test different parts of the construction: the RS reduction tests the REM--PPP evaluation, the annealed value tests the normalization of the correction field, the \(K=2\) and \(K=3\) reductions test the multilayer geometry, and the splitting-instability surface tests the local curvature of the functional.

\subsection*{Replica-symmetric reduction}

The vector replica-symmetric reduction corresponds to two effective blankets,
\[
\mathbf q_0=\mathbf 0,
\qquad
\mathbf q_1=\mathbf q,
\qquad
\mathbf q_2=\mathbf 1,
\]
\[
\zeta_0=0,
\qquad
\zeta_1=1,
\]
and terminal \(\zeta_2=1\). For layer \(p\), the variance increments in the terminal condition are
\[
(M\mathbf q)_p,
\qquad
(M(\mathbf 1-\mathbf q))_p.
\]
Thus we get the formula
\[
X_2^{(p)}
=
\log\cosh\left(
 h^{(p)}+
 \eta_{1,p}\sqrt{(M\mathbf q)_p}+
 \eta_{2,p}\sqrt{(M(\mathbf 1-\mathbf q))_p}
\right).
\]
Using, for \(Z\sim N(0,1)\), the exact identity
\[
\E_Z\cosh(a+bZ)=e^{b^2/2}\cosh(a),
\]
the step with \(\zeta_1=1\) gives
\[
X_1^{(p)}
=
\frac12(M(\mathbf 1-\mathbf q))_p
+
\log\cosh\left(
h^{(p)}+
\eta_{1,p}\sqrt{(M\mathbf q)_p}
\right).
\]
Since \(\zeta_0=0\), the first step is an ordinary Gaussian expectation:
\[
X_0^{(p)}
=
\frac12(M(\mathbf 1-\mathbf q))_p
+
\E_z\log\cosh\left(
h^{(p)}+
z\sqrt{(M\mathbf q)_p}
\right).
\]
Substituting this into \eqref{eq:final-functional-scalar}, the deterministic part is
\[
\frac12\sum_{p=1}^K\lambda_p (M(\mathbf 1-\mathbf q))_p
-
\frac12
\left(
\mathbf 1^\top M_1\mathbf 1
-
\mathbf q^\top M_1\mathbf q
\right).
\]
By applying the following identity
\[
\sum_{p=1}^K\lambda_p (Mv)_p
=
2\mathbf 1^\top M_1v,
\]
with \(v=\mathbf 1-\mathbf q\), this becomes
\[
\mathbf 1^\top M_1(\mathbf 1-\mathbf q)
-
\frac12
\left(
\mathbf 1^\top M_1\mathbf 1
-
\mathbf q^\top M_1\mathbf q
\right)
=
\frac12(\mathbf 1-\mathbf q)^\top M_1(\mathbf 1-\mathbf q).
\]
Hence the vector replica-symmetric expression is
\begin{equation}\label{eq:RS-scalar}
\cP_{\mathrm{RS}}(\mathbf q)
=
\log2+
\sum_{p=1}^K\lambda_p
\E_z\log\cosh\left(
h^{(p)}
+
z\sqrt{(M\mathbf q)_p}
\right)
+
\frac12
(\mathbf 1-\mathbf q)^\top M_1(\mathbf 1-\mathbf q).
\end{equation}
This is the multilayer RS reduction associated with the layer-wise CCM gauge.
The balanced scalar reduction is recovered by imposing the condition
\[
\mathbf q=q\mathbf 1.
\]
Then follow the relations
\[
(M\mathbf q)_p=d_pq,
\qquad
(\mathbf 1-\mathbf q)^\top M_1(\mathbf 1-\mathbf q)
=
\kappa(1-q)^2,
\]
and \eqref{eq:RS-scalar} becomes
\[
\cP_{\mathrm{RS}}(q)
=
\log2+
\sum_{p=1}^K\lambda_p
\E_z\log\cosh\left(h^{(p)}+z\sqrt{d_pq}\right)
+
\frac{\kappa}{2}(1-q)^2.
\]
This is the balanced-gauge specialization of the multilayer replica-symmetric expression for Gaussian-coupling DBM models \cite{AlbericiContucciMingione2021}.

As an annealed normalization check, set \(h^{(p)}=0\) and \(\mathbf q=\mathbf0\) in \eqref{eq:RS-scalar}. Then the Gaussian cavity term vanishes and
\[
\cP_{\mathrm{RS}}(\mathbf0)
=
\log2+
\frac12\mathbf1^\top M_1\mathbf1
=
\log2+
\sum_{p=1}^{K-1}\beta_p^2\lambda_p\lambda_{p+1}.
\]
This is the annealed high-temperature value of the adjacent-layer Gaussian RBM/DBM model, and therefore checks the normalization of the correction term in \eqref{eq:final-functional-scalar}.

\subsection*{RS splitting-instability surface}

The replica-symmetric vector reduction \eqref{eq:RS-scalar} also gives a local splitting-instability check for the CCM functional. Let \(\mathbf q\) be a self-consistent RS solution, satisfying
\[
q^{(p)}
=
\E_z\tanh^2\left(
h^{(p)}+
z\sqrt{(M\mathbf q)_p}
\right),
\qquad p=1,\ldots,K.
\]
Define the diagonal matrix
\[
\Gamma=\operatorname{diag}(\Gamma_1,\ldots,\Gamma_K),
\qquad
\Gamma_p=
\E_z\operatorname{sech}^4\left(
h^{(p)}+
z\sqrt{(M\mathbf q)_p}
\right).
\]
A first splitting of an RS blanket is governed, at linear order, by the multilayer stability matrix \(\Gamma M\). Thus the RS splitting-instability surface is
\begin{equation}\label{eq:RS-splitting-surface}
\rho(\Gamma M)=1,
\end{equation}
where \(\rho\) denotes the spectral radius. Equivalently, \(\Gamma M\) is similar to the symmetric tridiagonal matrix \(\mathcal K\) with \(\mathcal K_{p,p}=0\) and
\begin{equation}\label{eq:K-symmetric-instability}
\mathcal K_{p,p+1}
=
\mathcal K_{p+1,p}
=
2\beta_p^2
\sqrt{
\lambda_p\lambda_{p+1}
\Gamma_p\Gamma_{p+1}
},
\qquad p=1,\ldots,K-1.
\end{equation}
Therefore \eqref{eq:RS-splitting-surface} can also be written simply as
\[
\rho(\mathcal K)=1.
\]
The derivation of this condition from the one-split perturbation of the CCM functional is given in Appendix~\ref{app:RS-splitting-instability}.
For the two-layer RBM, write \(\beta_1=\beta\) and \(\mathbf q=(u,v)\). Then
\[
\Gamma_1=
\E_z\operatorname{sech}^4\left(
h^{(1)}+
z\sqrt{2\beta^2\lambda_2v}
\right),
\]
\[
\Gamma_2=
\E_z\operatorname{sech}^4\left(
h^{(2)}+
z\sqrt{2\beta^2\lambda_1u}
\right),
\]
and the instability condition becomes
\begin{equation}\label{eq:RBM-AT-like}
4\beta^4\lambda_1\lambda_2\Gamma_1\Gamma_2=1.
\end{equation}
In zero external field, on the paramagnetic RS branch \(u=v=0\), hence \(\Gamma_1=\Gamma_2=1\), and \eqref{eq:RBM-AT-like} reduces to the following closed expression:
\begin{equation}\label{eq:RBM-zero-field-critical}
4\beta^4\lambda_1\lambda_2=1.
\end{equation}
where \(\lambda_2=1-\lambda_1\). This reproduces the boundary of the annealed/high-temperature region of the bipartite spin glass of Barra--Genovese--Guerra~\cite{BarraGenoveseGuerra2011}. Thus the splitting-instability computation checks not only the RS value but also the local curvature of the CCM functional in the bipartite direction.

\subsection*{Two-layer RBM and symmetric three-layer DBM}

For \(K=2\), write \(\beta_1=\beta\) and the parameter is specialized as
\[
\mathbf q_\ell=(u_\ell,v_\ell).
\]
Then we can simplify the expressions for the cavity component
\[
(M(\mathbf q_\ell-\mathbf q_{\ell-1}))_1
=
2\beta^2\lambda_2(v_\ell-v_{\ell-1}),
\]
\[
(M(\mathbf q_\ell-\mathbf q_{\ell-1}))_2
=
2\beta^2\lambda_1(u_\ell-u_{\ell-1}),
\]
and then for the fugacity component
\[
\mathbf q_\ell^\top M_1\mathbf q_\ell
=
2\beta^2\lambda_1\lambda_2u_\ell v_\ell.
\]
The finite-level CCM functional becomes
\begin{equation}\label{eq:RBM-scalar-functional}
\cP_{\mathrm{RBM}}^{\mathrm{CCM}}(\zeta,u,v)
=
\log2+
\lambda_1X_0^{(1)}
+
\lambda_2X_0^{(2)}
-
\beta^2\lambda_1\lambda_2
\sum_{\ell=0}^{r-1}
\zeta_\ell
\left(
u_{\ell+1}v_{\ell+1}-u_\ell v_\ell
\right).
\end{equation}
This check is non-trivial because the two architectural layers carry distinct coordinates \(u_\ell\) and \(v_\ell\); the scalar diagonal specialization would not recover this bipartite structure in asymmetric cases.
The balanced RBM formula is recovered by setting \(u_\ell=v_\ell=q_\ell\).

Consider now \(K=3\) with parameters
\[
\lambda_1=\lambda_2=\lambda_3=\frac13,
\qquad
\beta_1=\beta_2=\beta.
\]
The size parameter is reduced to this setting as follows:
\[
\mathbf q_\ell=(a_\ell,b_\ell,c_\ell).
\]
If the two outer layers are treated symmetrically, we impose
\[
a_\ell=c_\ell=s_\ell,
\]
while the middle layer keeps its own coordinate \(b_\ell\). Then for the cavity component
\[
(M(\mathbf q_\ell-\mathbf q_{\ell-1}))_{\mathrm{out}}
=
\frac{2\beta^2}{3}(b_\ell-b_{\ell-1}),
\]
\[
(M(\mathbf q_\ell-\mathbf q_{\ell-1}))_{\mathrm{mid}}
=
\frac{4\beta^2}{3}(s_\ell-s_{\ell-1}),
\]
and for the fugacity component we have
\[
\mathbf q_\ell^\top M_1\mathbf q_\ell
=
\frac{4\beta^2}{9}s_\ell b_\ell.
\]
If the two outer layers have the same external field, the functional reduces to
\begin{equation}\label{eq:3L-scalar-functional}
\cP_{3\mathrm L}^{\mathrm{CCM}}(\zeta,s,b)
=
\log2+
\frac23X_0^{\mathrm{out}}
+
\frac13X_0^{\mathrm{mid}}
-
\frac{2\beta^2}{9}
\sum_{\ell=0}^{r-1}
\zeta_\ell
\left(
s_{\ell+1}b_{\ell+1}-s_\ell b_\ell
\right).
\end{equation}
Here \(X_0^{\mathrm{out}}\) and \(X_0^{\mathrm{mid}}\) are obtained from the same REM--PPP recursion with outer and middle variance scales displayed above. This check verifies that the middle layer, which is adjacent to two layers, receives twice the cavity scale of an outer layer in the symmetric three-layer model. The balanced three-layer formula is recovered by setting \(s_\ell=b_\ell=q_\ell\). At zero external field this gives
\[
\cP_{3\mathrm L,\mathrm{RS}}(q)
=
\log2+
\frac23\E_z\log\cosh\left(z\sqrt{\frac{2\beta^2}{3}q}\right)
+
\frac13\E_z\log\cosh\left(z\sqrt{\frac{4\beta^2}{3}q}\right)
+
\frac{2\beta^2}{9}(1-q)^2.
\]

\section{Nishimori-line deformation as an exact benchmark}\label{sec:nishimori-benchmark}

The model studied in the previous sections has centered Gaussian couplings but the CCM method is not limited to that.  As an exact benchmark for the layer-wise normalization, we can apply the same CCM procedure to the exactly solvable Nishimori deformation of the adjacent-layer DBM.  In the ordered-pair convention of Ref.~\cite{AlbericiCamilliContucciMingione2021}, the two orientations of each adjacent-layer edge carry independent couplings with tied mean and variance,
\[
\widetilde J_{ij}^{(p)}
\sim
N\left(
\frac{\mu_{p,p+1}}{2N},
\frac{\mu_{p,p+1}}{2N}
\right),
\]
and, when the random fields are included, the layer 
\(p\) field is on the same Nishimori line.  We remark that this is different from the centered model of the paper; the Nishimori deformation is used only as an exact functional benchmark for the CCM procedure.

The one-body Gaussian mark on the Nishimori line has equal mean and variance.  Thus a mark with scale \(A\) has the form \(A+\sqrt A\,z\), and its one-body contribution is
\begin{equation}\label{eq:Nishimori-psi}
\psi(A)
=
\E_z\log 2\cosh(A+\sqrt A\,z).
\end{equation}
In the layer-wise CCM variables, the scale in architectural layer \(L_p\) is \(h^{(p)}+(M\mathbf q)_p\).  The corresponding Nishimori deformation of the CCM functional is
\begin{equation}\label{eq:CCM-Nishimori-functional}
\cP_{\mathrm N}^{\mathrm{CCM}}(\mathbf q)
=
\sum_{p=1}^K
\lambda_p
\psi\left(h^{(p)}+(M\mathbf q)_p\right)
+
\frac12(\mathbf1-\mathbf q)^\top M_1(\mathbf1-\mathbf q)
-
\mathbf q^\top M_1\mathbf q.
\end{equation}
The first term is the Nishimori one-body cavity contribution.  The last term is the deterministic correction produced by the non-centered part of the Nishimori couplings.  The two changes with respect to the centered CCM functional---the one-body mark and the deterministic edge correction---are derived in Appendix~\ref{app:nishimori-deformation}.
With the identification
\begin{equation}\label{eq:Nishimori-identification}
\alpha_p^{\mathrm N}=\lambda_p,
\qquad
\mu_{p,p+1}=2\beta_p^2,
\end{equation}
the Nishimori interaction matrix has entries
\[
M_{p,p+1}^{\mathrm N}
=
\mu_{p,p+1}\alpha_{p+1}^{\mathrm N}
=
2\beta_p^2\lambda_{p+1}
=
M_{p,p+1},
\]
and the edge coefficient satisfies
\[
\frac{\Delta_{p,p+1}}2
=
\frac{\alpha_p^{\mathrm N}\mu_{p,p+1}\alpha_{p+1}^{\mathrm N}}2
=
\beta_p^2\lambda_p\lambda_{p+1}.
\]
Consequently, equation \eqref{eq:CCM-Nishimori-functional} coincides with the finite-dimensional variational integrand for the DBM on the Nishimori line~\cite{AlbericiCamilliContucciMingione2021}:
\begin{equation}\label{eq:Nishimori-integrand-match}
\cP_{\mathrm N}^{\mathrm{CCM}}(\mathbf q)
=
p_{\mathrm{var}}(\mathbf q;\mu,h).
\end{equation}
This is an important benchmark.  It checks the vector structure of the CCM functional, including the matrices \(M=2M_0\Lambda\), \(M_1=\Lambda M_0\Lambda\), and the layer-wise coordinate \(\mathbf q\).  It does not give a new proof of the Nishimori min--max selection rule: once the two integrands are identified, the known selection rule is inherited.

\section{On the local selection problem}\label{sec:local-selection}

The procedure given above is to obtain the finite-level CCM functional. The problem of selecting the physical value of this functional is a separate variational and stability problem. We do not prove here the global theorem selecting the thermodynamic value. This will be explored in a separate paper. We only recall the usual local criterion naturally associated with the functional.

For a fixed number of blankets, a candidate state is a point
\[
(\zeta,\mathbf q)
\]
in the admissible domain of \eqref{eq:final-functional-scalar}. If all the relevant coordinates are internal, the local selection equations are the stationarity conditions
\[
\nabla_{\mathbf q_\ell}\cP_{\mathrm{RBM/DBM}}^{\mathrm{CCM}}(\zeta,\mathbf q)=0,
\qquad
\partial_{\zeta_\ell}\cP_{\mathrm{RBM/DBM}}^{\mathrm{CCM}}(\zeta,\mathbf q)=0.
\]
When a rate is on the boundary, or two consecutive blankets merge, these equations are replaced by the corresponding Karush-Kuhn-Tucker (KKT) conditions. These conditions select stationary candidates in the blanket coordinates and Poisson exponents.

Stationarity alone is not sufficient. A locally selected state must also be stable with respect to further splitting of any active blanket. For a self-consistent RS solution, this stability test produces the matrix condition already displayed in \eqref{eq:RS-splitting-surface},
\[
\rho(\Gamma M)\le 1.
\]
If the spectral radius is larger than one, then the corresponding blanket is locally unstable and has to be refined. The same logic applies after one splitting: the stability matrix is obtained by replacing the RS susceptibility by the susceptibility computed in the tilted one-split state.

We remark that this is not a global selection theorem. It is the CCM version of the standard finite-step stationarity and stability analysis used in spin-glass variational functionals, including the SK cavity/Parisi setting and its minimization theory \cite{AizenmanSimsStarr2003,Guerra2003,Talagrand2006,AuffingerChen2015}, the de Almeida--Thouless stability analysis \cite{deAlmeidaThouless1978}, and bipartite or multilayer mean-field variational structures \cite{BarraGenoveseGuerra2011,Genovese2023,AlbericiCamilliContucciMingione2021,AlbericiContucciMingione2021}. In particular, whenever the CCM functional reduces exactly to a known variational integrand, the known selection rule is inherited automatically. The equality of the functionals is the actual consistency check; the selection rule is then a consequence of the known theory.

The usefulness of the local criterion is therefore diagnostic. Away from regimes where a global selection theorem is available, one can study candidate physical states by solving the finite-level stationarity equations, imposing KKT conditions on boundary rates or merged blankets, and recursively refining any blanket whose splitting matrix is unstable. This procedure does not change the derivation of the functional; it is a local method for exploring its possible physical selection.

\section{Relation with other full-RSB formulations}\label{sec:full-rsb-comparisons}

We compare the functional \eqref{eq:final-functional-scalar} with the bipartite replica calculation of Hartnett, Parker and Geist~\cite{HartnettParkerGeist2018} and the multi-species formulation of Chen, Issa and Mourrat~\cite{ChenIssaMourrat2026}. Both comparisons concern centered Gaussian couplings and zero external fields.

Write the finite hierarchy as the nondecreasing step function
\begin{equation}\label{eq:full-rsb-step-path}
\mathbf q(x)=\mathbf q_\ell,
\qquad
\zeta_{\ell-1}\leq x<\zeta_\ell,
\qquad \ell=0,\ldots,r.
\end{equation}
The endpoint conventions are those of Section~\ref{sec:ppp}. Summation by parts rewrites the quadratic correction as
\begin{equation}\label{eq:full-rsb-summation-by-parts}
\int_0^1\mathbf q(x)^\top M_1\mathbf q(x)\,dx
=
\kappa-
\sum_{\ell=0}^{r-1}\zeta_\ell
\left(
\mathbf q_{\ell+1}^\top M_1\mathbf q_{\ell+1}
-
\mathbf q_\ell^\top M_1\mathbf q_\ell
\right).
\end{equation}

\subsection*{Bipartite model}

For the specialization \eqref{eq:RBM-scalar-functional}, choose unit coupling scale in Ref.~\cite{HartnettParkerGeist2018} and denote its inverse temperature by $\beta_{\mathrm H}$. The parameter correspondence is
\begin{equation}\label{eq:hartnett-parameter-correspondence}
\begin{gathered}
\alpha_v=\lambda_1,\qquad
\alpha_h=\lambda_2,\qquad
\beta_{\mathrm H}=\sqrt2\,\beta(\lambda_1\lambda_2)^{1/4},
\\
q^{(v)}(x)=u(x),\qquad q^{(h)}(x)=v(x).
\end{gathered}
\end{equation}
This matches their coupling normalization, given before Eq.~(2.7), with the Hamiltonian \eqref{eq:hamiltonian}. Their variance parameters, defined in Eq.~(2.43), then satisfy
\begin{equation}\label{eq:hartnett-cavity-variances}
\beta_{\mathrm H}^{\,2}r^{(h)}(x)=2\beta^2\lambda_2v(x),
\qquad
\beta_{\mathrm H}^{\,2}r^{(v)}(x)=2\beta^2\lambda_1u(x).
\end{equation}
These are precisely the two cavity variances of our bipartite construction. Their Parisi equation (2.44), written in dimensionless field coordinates, therefore gives the Gaussian recursion \eqref{eq:X-recursion-scalar}.

The unit-exponent Gaussian step completes each terminal variance to its diagonal value. Combining this contribution with \eqref{eq:full-rsb-summation-by-parts} reproduces the endpoint terms and the integral of the product of the two profiles in their Eq.~(2.42). Thus \eqref{eq:RBM-scalar-functional} coincides with their full-RSB expression in pressure units: the pressure is minus their inverse temperature times their free-energy density. The identification holds for each trial hierarchy, before imposing stationarity.

\subsection*{General multilayer model}

Chen, Issa and Mourrat~\cite{ChenIssaMourrat2026} normalize the species overlaps by the total number of spins. The corresponding path and covariance polynomial for our model are
\begin{equation}\label{eq:chen-multispecies-correspondence}
\mathbf p(x)=\Lambda\mathbf q(x),
\qquad
\Xi(\mathbf a)=\mathbf a^\top M_0\mathbf a.
\end{equation}
Write $\Psi$ for their species-weighted, compensated one-site cascade functional, defined in Eqs.~(2.1)--(2.7) and~(3.7). At time one half and zero auxiliary field, their Parisi functional (3.14) specializes to
\begin{equation}\label{eq:chen-parisi-specialization}
\mathcal P_{1/2,0}[\Lambda\mathbf q]
=
\Psi\!\left[\frac12M\mathbf q\right]
-\frac12\int_0^1\mathbf q(x)^\top M_1\mathbf q(x)\,dx.
\end{equation}
The factor one half matches their Gaussian normalization with the Hamiltonian \eqref{eq:hamiltonian}.

On the step hierarchy, their one-site calculation uses the same recursion \eqref{eq:X-recursion-scalar}. Including its deterministic compensation and the unit-exponent terminal step gives
\begin{equation}\label{eq:chen-single-site-compensation}
\Psi\!\left[\frac12M\mathbf q\right]
=
\frac12\sum_{p=1}^K\lambda_p(M\mathbf1)_p
-\sum_{p=1}^K\lambda_pX_0^{(p)}
=
\kappa-\sum_{p=1}^K\lambda_pX_0^{(p)}.
\end{equation}
Substituting \eqref{eq:full-rsb-summation-by-parts} and \eqref{eq:chen-single-site-compensation} into \eqref{eq:chen-parisi-specialization} yields
\begin{equation}\label{eq:chen-ccm-functional-identity}
\cP_{\mathrm{RBM/DBM}}^{\mathrm{CCM}}(\zeta,\mathbf q)
=
\log2+\frac{\kappa}{2}
-\mathcal P_{1/2,0}[\Lambda\mathbf q].
\end{equation}
Thus the complete functional \eqref{eq:final-functional-scalar} is also obtained by specializing the multi-species Parisi functional. The identity holds at every finite depth and extends to bounded monotone paths by the continuity established in Lemma~3.6 of Ref.~\cite{ChenIssaMourrat2026}. Also their sup--inf characterization of the thermodynamic value at zero deterministic field given in Theorem~1.1 is consistent with our local selection approach given before. These identifications connect the blanket calculation with the replica and Hamilton--Jacobi formulations.

\section{Why CCM works}\label{sec:why-ccm-works}

This section is independent from the derivation of the RBM/DBM functional above and may be skipped by readers interested only in the explicit formula and its checks. 
The CCM~\cite{SimplifiedParisiIII} originates in the Kernel/Blanket theory of disordered systems~\cite{RSBWithoutReplicas,UniversalScaling,SimplifiedParisi}. The blanket-theoretic viewpoint is related to the Free Energy Principle and the Markov-blanket formulation of variational free energy~\cite{Friston2010FEP,Friston2013Life,Kirchhoff2018MarkovBlankets} and it can be interpreted as a purely kinetic LFT of the kind described in \cite{FranchiniBardella2026,BardellaFranchini2024,BardellaFranchini2025} by identifying the potential energy with the vanishing core contribution.
Here we connect the calculations of Section~\ref{sec:io-balanced}--\ref{sec:ppp} with this viewpoint~\cite{SimplifiedParisiIII} in the low temperature limit. 

The blanket decomposition separates interactions within each blanket from those connecting it to preceding blankets. Since each blanket crosses the architectural layers, its core contains couplings between cells in adjacent layers. Writing the sum of the interface energies as \(\overline H_N\), the full and IO partition functions are
\begin{equation}\label{eq:dbm-partition-functions}
Z_N=\sum_\sigma e^{H_N(\sigma)},\qquad
\overline Z_N=\sum_\sigma e^{\overline H_N(\sigma)}.
\end{equation}
Gaussian interpolation of the independent core couplings gives
\begin{equation}\label{eq:dbm-core-pressure-bound}
0\leq\frac1N\E\log\frac{Z_N}{\overline Z_N}
\leq\frac1{N^2}\sum_{p=1}^{K-1}\beta_p^2
\sum_{\ell=1}^{r}|V_\ell\cap L_p|\,|V_\ell\cap L_{p+1}|.
\end{equation}
The cell sizes in Eq.~\eqref{eq:balanced-block} make this bound vanish as the largest blanket fraction in each architectural layer tends to zero. Taking the thermodynamic limit at fixed partition and then refining it therefore reduces the pressure to that of the IO model~\cite[Section~2.1]{SimplifiedParisiIII}, while the correction field retains the complete edge-shell counts of Eq.~\eqref{eq:y-level-var-scalar}. The resulting interface is linear in the spins of the new blanket:
\begin{equation}\label{eq:dbm-direct-interface}
\overline H_\ell(\sigma)
=\sum_{p=1}^{K}\sum_{i\in V_\ell\cap L_p}
\sigma_i^{(p)}\overline h_{\ell,i}^{(p)}(\sigma_{U_{\ell-1}}),
\end{equation}
\begin{equation}\label{eq:dbm-direct-field}
\overline h_{\ell,i}^{(p)}(\sigma_{U_{\ell-1}})
=\sqrt{\frac2N}\,\beta_{p-1}\sum_{j\in U_{\ell-1}\cap L_{p-1}}
J_{ji}^{(p-1)}\sigma_j^{(p-1)}
+\sqrt{\frac2N}\,\beta_p\sum_{k\in U_{\ell-1}\cap L_{p+1}}
J_{ik}^{(p)}\sigma_k^{(p+1)}.
\end{equation}
Terms involving absent architectural layers are omitted.

This IO model provides the reference for the low-temperature description. We rectify configurations with respect to a global ground state, which maximizes the Hamiltonian in our convention. Energies and Gibbs weights are transported together, and the overlaps are preserved~\cite[Sections~2.2--2.4]{SimplifiedParisiIII}. Excitations then form a dilute gas of flipped spins whose complete energy costs include interactions with preceding and subsequent blankets and between defects. For finitely many nodes with finitely many defects, independently sampled supports become disjoint with probability tending to one as the cells grow; defect interactions vanish, and the new energy contributions are jointly approximated at fixed disorder by independent sums of complete local costs. Descendants retain their ancestors' defects and energies, while new nodes receive new arrangements~\cite[Sections~3.1--3.2]{SimplifiedParisiIII}. Two states thus share both spins and energy contributions up to their last common ancestor, which is the origin of the hierarchical structure. Because each node specifies all architectural cells of its blanket, this genealogy determines the overlaps in every species.

To obtain thermodynamic weights, we extend the finite construction to exponential populations within the magnetization shells. At fixed small positive defect densities, these populations can be chosen with small relative intersections throughout every kernel strip, including between nodes with different parents; the thermodynamic limit is taken at fixed partition and followed by dilution~\cite[Section~3.3]{SimplifiedParisiIII}. Along each branch, the complete centered energy is built from centered increments that are linear in the new defects once the prefix is fixed. Their coefficients contain the preceding configuration and the mean fields of subsequent blankets, so sampling the defects leads to the local problem of REM universality~\cite{SimplifiedParisiII,REMUniversalityLinear,REMUniversalityLetter}. For the coefficient populations selected by the ground state, the regularity needed in this step is physically justified in Ref.~\cite[Section~3.3]{SimplifiedParisiIII}. We then adopt the extensive GREM approximation, representing the conditional increment laws by a common law at each level, independently realized at distinct nodes. Comparing the rarity of energies with the number of available states determines which consecutive levels condense together and fixes their Poisson exponents~\cite[Section~3.4]{SimplifiedParisiIII}. In assembling the resulting Gibbs weights, we retain the energy centers and shell entropies and include all descendant continuations in the mass of each prefix~\cite[Section~2.4]{SimplifiedParisiIII}.

This construction supplies the Gibbs masses and species overlaps on the same states. The overlaps in turn determine the complete law of the fresh cavity fields: conditioned on the reservoir, these are jointly Gaussian, and rectification preserves their law because their couplings are independent of the disorder selecting the reference~\cite[Section~4.2]{SimplifiedParisiIII}. For two configurations indexed by the paths \(\alpha\) and \(\gamma\), averaging over the fresh noise gives
\begin{equation}\label{eq:dbm-linear-cavity-covariance}
\E\!\left[x_{N,p}(\sigma^\alpha)x_{N,p}(\sigma^\gamma)\right]
=\frac{2\beta_{p-1}^{2}}N\sum_{i\in L_{p-1}}\sigma_i^{(p-1),\alpha}\sigma_i^{(p-1),\gamma}
+\frac{2\beta_p^{2}}N\sum_{j\in L_{p+1}}\sigma_j^{(p+1),\alpha}\sigma_j^{(p+1),\gamma},
\end{equation}
\begin{equation}\label{eq:dbm-correction-cavity-covariance}
\E\!\left[y_N(\sigma^\alpha)y_N(\sigma^\gamma)\right]
=\frac2{N^2}\sum_{p=1}^{K-1}\beta_p^2
\sum_{i\in L_p}\sigma_i^{(p),\alpha}\sigma_i^{(p),\gamma}
\sum_{j\in L_{p+1}}\sigma_j^{(p+1),\alpha}\sigma_j^{(p+1),\gamma}.
\end{equation}
Normalizing the spin sums by the respective layer sizes identifies the species overlaps. In the thermodynamic limit, the two covariances give the action of \(M\) and the quadratic form defined by \(M_1\); together with the independence of the fresh-noise families, they fix the joint Gaussian law. The physical overlaps include the magnetization contributions from cells beyond the shared prefix, whose spins remain aligned on average with the ground state~\cite[Section~4.3]{SimplifiedParisiIII}. These overlaps increase componentwise to one.

Following the operational correspondence established in Ref.~\cite[Section~4.7]{SimplifiedParisiIII}, we choose the constructive cell fractions in each architectural layer to represent their common part and successive increments, retaining the same state masses, ancestry, and averaging exponents. With these fractions, the amplitudes in Eqs.~\eqref{eq:xp-constructive-scalar} and~\eqref{eq:y-constructive-scalar} reproduce the full covariance increments, so the masses and complete cavity fields have the same joint law as in the physical representation. The common linear fields remain inside the hyperbolic cosines throughout the recursive averages and are integrated last at zero exponent. The common correction field remains in the random logarithmic contribution and has zero mean over the fresh noise~\cite[Sections~4.4--4.6]{SimplifiedParisiIII}.

Using these complete fields with the same Gibbs masses and ancestors identifies all cavity contributions jointly with the ROSt. The cascade can be reconstructed from one independent PPP per level by assigning a parent to each node and normalizing the terminal masses globally. Descendants inherit the assignments and Gaussian increments of their ancestors. Transporting the complete fields with these same assignments preserves the random cavity functional in joint distribution~\cite[Sections~3.5--3.6 and~4.5]{SimplifiedParisiIII}. If several physical fibres contribute, their weighted contributions are combined before taking the logarithms, as in Section~\ref{sec:ass-substitution}. The mean logarithms are then evaluated by Eqs.~\eqref{eq:X-recursion-scalar} and~\eqref{eq:correction-ppp-value-scalar}, yielding Eq.~\eqref{eq:final-functional-scalar}~\cite[Section~4.6]{SimplifiedParisiIII}. The parameters remain free during this calculation and are selected as discussed in Sections~\ref{sec:local-selection} and~\ref{sec:full-rsb-comparisons}. An extension of this low temperature argument to arbitrary temperature is possible through the REM universality approach developed in Refs.~\cite{REMUniversalityLinear,REMUniversalityLetter}.

\section{Discussion}\label{sec:discussion}

The calculation above applies the CCM of \cite{SimplifiedParisiIII} to dense Ising RBM/DBM models with Gaussian couplings in a layer-wise constructive gauge. The blanket \(\ell\) vector of parameter variables
\[
\mathbf q_\ell=(q_\ell^{(1)},\ldots,q_\ell^{(K)})
\]
is the vector of cumulative layer fractions in the constructive partition,
\[
|U_\ell\cap L_p|=\lambda_pq_\ell^{(p)}N+o(N),
\]
and supplies the levels of the nested REM--PPP evaluation of the cavity functional.
The model-specific content of the computation is contained in two matrix quantities. The cavity field in layer \(p\) has scale
\[
Q_\ell^{(p)}=(M\mathbf q_\ell)_p,
\qquad
M=2M_0\Lambda,
\]
while the correction field has scale
\[
Q_\ell=\mathbf q_\ell^\top M_1\mathbf q_\ell,
\qquad
M_1=\Lambda M_0\Lambda.
\]
These amplitudes enter the blanket components \eqref{eq:xp-constructive-scalar} and \eqref{eq:y-constructive-scalar}. CCM identifies the blanket decomposition of the cavity variables with the Aizenman--Sims--Starr ROSt, and nested REM--PPP averaging gives the functional~\cite{SimplifiedParisiIII}.
Notice that the balanced gauge is the diagonal specialization
\[
\mathbf q_\ell=q_\ell\mathbf1.
\]
In this specialization one recovers
\[
Q_\ell^{(p)}=d_pq_\ell,
\qquad
Q_\ell=\kappa q_\ell^2,
\]
with the parameter \(d_p\) and \(\kappa\) as follows:
\[
d_p=(M\mathbf1)_p,
\qquad
\kappa=\mathbf1^\top M_1\mathbf1.
\]
Thus the scalar balanced formula is the diagonal case of the layer-wise functional.

The final result is the finite-level functional of \eqref{eq:final-functional-scalar}. For \(K=2\), it gives the dense bipartite Ising RBM with Gaussian couplings and for \(K\ge3\) it gives the corresponding adjacent-layer DBM. Its vector replica-symmetric reduction agrees with the expected multilayer RS structure, while the balanced RS expression is recovered by imposing \(\mathbf q=q\mathbf1\).
Finally, the RS splitting-instability check tests the local curvature of the functional: in the zero-field bipartite specialization it yields \(4\beta^4\lambda_1\lambda_2=1\), matching the known annealed boundary of the bipartite spin glass~\cite{BarraGenoveseGuerra2011}.

The Nishimori-line deformation gives another exact benchmark: under \(\alpha_p^{\mathrm N}=\lambda_p\) and \(\mu_{p,p+1}=2\beta_p^2\), the CCM functional reduces to the known Nishimori variational integrand, confirming the vector layer-wise normalization of \(M\) and \(M_1\)~\cite{AlbericiCamilliContucciMingione2021}.

Section~\ref{sec:full-rsb-comparisons} connects the full functional with two independent formulations. For centered Gaussian couplings at zero external field, the bipartite specialization coincides with the replica expression of Hartnett, Parker and Geist~\cite{HartnettParkerGeist2018}, while \eqref{eq:chen-ccm-functional-identity} identifies the general multilayer functional with that of Chen, Issa and Mourrat~\cite{ChenIssaMourrat2026}. The CCM thus provides a constructive derivation of these full-RSB expressions through blanket amplitudes and REM universality. In this setting, the thermodynamic value is characterized by the sup--inf theorem of Ref.~\cite{ChenIssaMourrat2026}, while Section~\ref{sec:local-selection} studies local stationarity and splitting stability at finite depth.

The same counting principle suggests natural extensions. If the layer graph is not a path, \(M_0\) is replaced by the weighted adjacency matrix of that graph, and \(M=2M_0\Lambda\), \(M_1=\Lambda M_0\Lambda\) are defined as before. For higher-order interactions, rectangular edge shells would be replaced by the corresponding hyperedge shells, with the appropriate combinatorial factors. The practical CCM rule remains the same: decompose the cavity variables into blanket components, identify their ROSt realization, and perform the nested REM--PPP average with the model-specific amplitudes.

\appendix

\section{Technical details of the layer-wise CCM computation}\label{app:technical}

This appendix records the normalizations and finite-level identities used in the layer-wise scalar cavity functional. It is not needed for the flow of the main text, but it makes explicit where the factors in Sections~\ref{sec:constructive-variables} and~\ref{sec:ppp} come from.
We use the boundary convention
\[
\beta_0=\beta_K=0,
\qquad
\lambda_0=\lambda_{K+1}=0.
\]
With this convention, the balanced specialization gives
\[
d_p=2\bigl(\beta_{p-1}^2\lambda_{p-1}+\beta_p^2\lambda_{p+1}\bigr)
\]
for all layers \(p=1,\ldots,K\). Also,
\[
\kappa
=2\sum_{p=1}^{K-1}\beta_p^2\lambda_p\lambda_{p+1}
=\mathbf 1^\top M_1\mathbf 1.
\]
Indeed, since \(M_1=\Lambda M_0\Lambda\) and \(M_0\) is symmetric with off-diagonal entries \(\beta_p^2\),
\[
\mathbf 1^\top M_1\mathbf 1
=
\sum_{a,b=1}^K \lambda_a(M_0)_{ab}\lambda_b
=
2\sum_{p=1}^{K-1}\beta_p^2\lambda_p\lambda_{p+1}.
\]
The factor \(2\) appears because the symmetric quadratic form counts both ordered pairs \((p,p+1)\) and \((p+1,p)\).
For the cavity field, the blanket-\(\ell\) contribution in layer \(p\) is
\[
x_{N,p,\ell}(\sigma)
=
\frac{\sqrt2}{\sqrt N}
\left(
\beta_{p-1}
\sum_{i\in V_\ell\cap L_{p-1}}
\widetilde J_i^{(p-1)}\sigma_i^{(p-1)}
+
\beta_p
\sum_{j\in V_\ell\cap L_{p+1}}
\widehat J_j^{(p)}\sigma_j^{(p+1)}
\right).
\]
Conditioning on the reservoir configuration, the cross terms vanish because the fresh couplings are independent and centered. Since the spins are Ising variables,
\[
\Var(x_{N,p,\ell}\mid\sigma)
=
\frac{2\beta_{p-1}^2}{N}|V_\ell\cap L_{p-1}|
+
\frac{2\beta_p^2}{N}|V_\ell\cap L_{p+1}|.
\]
Using the layer-wise estimate
\[
|V_\ell\cap L_a|
=
\lambda_a\bigl(q_\ell^{(a)}-q_{\ell-1}^{(a)}\bigr)N+o(N)
\]
gives the formula for the variance:
\[
\Var(x_{N,p,\ell}\mid\sigma)
=
\bigl(M(\mathbf q_\ell-\mathbf q_{\ell-1})\bigr)_p+o(1).
\]
At a fixed blanket, the neighboring-layer sums are independent Gaussians, so their variances add. This fixes the amplitude of the prefix-indexed cavity variable of Section~\ref{sec:constructive-variables}:
\[
x_p^\alpha
=
\sum_{\ell=1}^{r}
z_{\ell,p}(\alpha_{\leq\ell})
\sqrt{
\bigl(M(\mathbf q_\ell-\mathbf q_{\ell-1})\bigr)_p
}.
\]
For the correction field, the blanket-\(\ell\) shell between two adjacent layers is
\[
W_\ell^{(p,p+1)}
=
B_\ell^{(p,p+1)}\setminus B_{\ell-1}^{(p,p+1)},
\qquad
\ell=1,\ldots,r,
\] 
\[
B_\ell^{(p,p+1)}=(U_\ell\cap L_p)\times(U_\ell\cap L_{p+1}).
\]
Thus \(B_0^{(p,p+1)}=\varnothing\) and \(B_r^{(p,p+1)}=L_p\times L_{p+1}\). The layer-wise gauge gives
\[
|W_\ell^{(p,p+1)}|
=
\lambda_p\lambda_{p+1}N^2
\left(
q_\ell^{(p)}q_\ell^{(p+1)}
-
q_{\ell-1}^{(p)}q_{\ell-1}^{(p+1)}
\right)
+o(N^2).
\]
Let us introduce a symbol for the difference 
\[
\Delta q_\ell^{(p)}=q_\ell^{(p)}-q_{\ell-1}^{(p)},
\]
then the increment decomposes as
\[
q_\ell^{(p)}q_\ell^{(p+1)}
-
q_{\ell-1}^{(p)}q_{\ell-1}^{(p+1)}
=
q_{\ell-1}^{(p)}\Delta q_\ell^{(p+1)}
+
q_{\ell-1}^{(p+1)}\Delta q_\ell^{(p)}
+
\Delta q_\ell^{(p)}\Delta q_\ell^{(p+1)}.
\]
The last term is the finite-level core contribution controlled on refinement in the IO reduction. The complete finite shell gives the correction variance below.
The correction mark associated with this shell is
\[
y_{N,\ell}^{(p,p+1)}(\sigma)
=
\frac{\sqrt2\,\beta_p}{N}
\sum_{(i,j)\in W_\ell^{(p,p+1)}}
\overline J_{ij}^{(p)}
\sigma_i^{(p)}\sigma_j^{(p+1)}.
\]
The \(1/N\) normalization keeps the variance of this \(O(N^2)\) edge sum of order one. Again, cross terms vanish by independence and centering, and each diagonal term contributes one because the spins are Ising. Hence we get
\[
\Var\bigl(y_{N,\ell}^{(p,p+1)}\mid\sigma\bigr)
=
\frac{2\beta_p^2}{N^2}|W_\ell^{(p,p+1)}|.
\]
Therefore, the variance is rewritten as follows
\[
\Var\bigl(y_{N,\ell}^{(p,p+1)}\mid\sigma\bigr)
=
2\beta_p^2\lambda_p\lambda_{p+1}
\left(
q_\ell^{(p)}q_\ell^{(p+1)}
-
q_{\ell-1}^{(p)}q_{\ell-1}^{(p+1)}
\right)
+o(1).
\]
The contributions of distinct edge-shells use independent fresh couplings, so the variances add:
\[
Q_\ell-Q_{\ell-1}
=
\mathbf q_\ell^\top M_1\mathbf q_\ell
-
\mathbf q_{\ell-1}^\top M_1\mathbf q_{\ell-1},
\qquad
Q_\ell=\mathbf q_\ell^\top M_1\mathbf q_\ell.
\]
The \(o(N)\) and \(o(N^2)\) errors in the block and shell cardinalities produce only \(o(1)\) errors in the corresponding variances.

The Gaussian step of the nested REM--PPP average in Section~\ref{sec:ppp} is elementary. If \(g\) is a standard Gaussian and \(v\ge0\), then the identity holds
\[
\frac1m\log\E\exp(mg\sqrt v)
=
\frac{m}{2}v
\]
for \(m>0\), the value is \(0\) for \(m=0\) by continuity. Applying this identity blanket by blanket to
\[
y^\alpha
=
\sum_{k=1}^{r}
g_k(\alpha_{\leq k})
\sqrt{
\mathbf q_k^\top M_1\mathbf q_k
-
\mathbf q_{k-1}^\top M_1\mathbf q_{k-1}
}
\]
gives the expression for the fugacity correction
\[
\E\log\langle e^{y^\alpha}\rangle_\xi
=
\frac12
\sum_{\ell=0}^{r-1}
\zeta_\ell
\left(
\mathbf q_{\ell+1}^\top M_1\mathbf q_{\ell+1}
-
\mathbf q_\ell^\top M_1\mathbf q_\ell
\right).
\]
This is the finite-level form of the CCM interface reduction used in the main text. The balanced scalar expression is recovered by setting \(\mathbf q_\ell=q_\ell\mathbf1\).

\section{Finite-size identities for the symmetric three-layer model}\label{app:numerics}

This appendix records the finite-size identities for the symmetric three-layer model. They are diagnostic identities only; they are not used in the derivation of the layer-wise CCM functional. The point is that, for moderate layer size, the partition function can be summed exactly over the middle layer.
Take
\[
K=3,
\qquad
N_1=N_2=N_3=n,
\qquad
N=3n,
\qquad
\beta_1=\beta_2=\beta,
\qquad
h=0.
\]
Let $J^{(1)}$ and $J^{(2)}$ have independent standard Gaussian entries. Writing the middle-layer spin as $s\in\{\pm1\}^n$, the Hamiltonian is
\[
H_N(\sigma^{(1)},s,\sigma^{(3)})
=
\frac{\sqrt2\beta}{\sqrt{3n}}
\left(
\sum_{i,j=1}^nJ_{ij}^{(1)}\sigma_i^{(1)}s_j
+
\sum_{j,k=1}^nJ_{jk}^{(2)}s_j\sigma_k^{(3)}
\right).
\]
For fixed $s$, the two outer layers can be summed independently. With
\[
c=\frac{\sqrt2\beta}{\sqrt{3n}},
\]
one obtains the formula
\[
Z_N(\beta)
=
\sum_{s\in\{\pm1\}^n}
\prod_{i=1}^n2\cosh\bigl(c(J^{(1)}s)_i\bigr)
\prod_{k=1}^n2\cosh\bigl(c((J^{(2)})^\top s)_k\bigr).
\]
Thus the formula for the average pressure
\[
p_N(\beta)=\frac1N\E_J\log Z_N(\beta)
\]
can be estimated by exact spin summation for each disorder sample, followed by an empirical average over the disorder.

The same reduction gives an exact finite-size ground-state identity. For fixed $s$, the optimal outer spins align with their local fields, hence
\[
\max_{\sigma^{(1)},\sigma^{(3)}}
H_N(\sigma^{(1)},s,\sigma^{(3)})
=
\frac{\sqrt2\beta}{\sqrt{3n}}
\left(
\|J^{(1)}s\|_1+
\|(J^{(2)})^\top s\|_1
\right).
\]
Therefore,
\[
e_N^{(\beta)}(J)
=
\frac1N\max_\sigma H_N(\sigma)
=
\frac{\sqrt2\beta}{(3n)^{3/2}}
\max_{s\in\{\pm1\}^n}
\left(
\|J^{(1)}s\|_1+
\|(J^{(2)})^\top s\|_1
\right).
\]
Ground-state values quoted per unit of $\beta$ correspond to setting $\beta=1$ in this expression.
Pair observables can also be sampled without Markov-chain Monte Carlo once the middle layer has been enumerated. The exact Gibbs weight of the middle configuration is proportional to
\[
w(s)
\propto
\prod_{i=1}^n2\cosh\bigl(c(J^{(1)}s)_i\bigr)
\prod_{k=1}^n2\cosh\bigl(c((J^{(2)})^\top s)_k\bigr).
\]
Conditioned on $s$, the outer spins are independent with means
\[
m_i^{(1)}(s)=\tanh\bigl(c(J^{(1)}s)_i\bigr),
\qquad
m_k^{(3)}(s)=\tanh\bigl(c((J^{(2)})^\top s)_k\bigr).
\]
These identities are useful for checking numerical implementations of the layer-wise functional and of its balanced scalar specialization. They do not determine the physical selection rule in the non-convex regime.

\section{RS splitting-instability calculation}\label{app:RS-splitting-instability}

This appendix shows the linear calculation behind the RS splitting-instability surface used in Section~\ref{sec:checks}. The calculation is local around a vector RS solution of the CCM functional and tests whether an RS blanket can be split into two nearby blankets.
Let \(\mathbf q\) be a self-consistent RS solution,
\[
q^{(p)}
=
\E_z\tanh^2\left(
h^{(p)}+z\sqrt{(M\mathbf q)_p}
\right),
\qquad p=1,\ldots,K.
\]
Consider a one-split perturbation of the RS blanket. The two nearby blanket coordinates are denoted by \(\mathbf q^{(0)}\) and \(\mathbf q^{(1)}\), and their difference is
\[
\boldsymbol\delta
=
\mathbf q^{(1)}-\mathbf q^{(0)},
\qquad
\|\boldsymbol\delta\|\ll1.
\]
The corresponding infinitesimal increment of the cavity variance in layer \(p\) is
\[
\Delta A_p=(M\boldsymbol\delta)_p.
\]
The one-split self-consistency equations say that the difference between the two nearby blanket coordinates is, to first order, the local variance of the magnetization response produced by the additional Gaussian increment. Thus
\[
\delta^{(p)}
=
\E_z\operatorname{sech}^4\left(
h^{(p)}+z\sqrt{(M\mathbf q)_p}
\right)
(M\boldsymbol\delta)_p
+o(\|\boldsymbol\delta\|).
\]
Define the function
\[
\Gamma_p
=
\E_z\operatorname{sech}^4\left(
h^{(p)}+z\sqrt{(M\mathbf q)_p}
\right),
\qquad
\Gamma=\operatorname{diag}(\Gamma_1,\ldots,\Gamma_K).
\]
Then the linearized splitting equation is
\begin{equation}\label{eq:appendix-splitting-linear}
\boldsymbol\delta
=
\Gamma M\boldsymbol\delta.
\end{equation}
The boundary of linear stability of the RS solution is therefore
\begin{equation}\label{eq:appendix-splitting-spectral}
\rho(\Gamma M)=1.
\end{equation}
This gives the RS splitting-instability surface.
It is useful to rewrite the same condition with a symmetric matrix. Assume first that \(\Gamma_p>0\) on the layers under consideration and put
\[
D=\operatorname{diag}\left(\sqrt{\lambda_1/\Gamma_1},\ldots,\sqrt{\lambda_K/\Gamma_K}\right).
\]
Then, the matrix \(\mathcal K\) is defined as follows:
\[
\mathcal K=D\Gamma M D^{-1},
\]
it is symmetric and tridiagonal. Its diagonal entries vanish, and its adjacent-layer entries are
\begin{equation}\label{eq:appendix-K-entries}
\mathcal K_{p,p+1}
=
\mathcal K_{p+1,p}
=
2\beta_p^2
\sqrt{
\lambda_p\lambda_{p+1}
\Gamma_p\Gamma_{p+1}
},
\qquad p=1,\ldots,K-1.
\end{equation}
Since \(\mathcal K\) is similar to \(\Gamma M\), we have
\[
\rho(\Gamma M)=\rho(\mathcal K),
\]
and the instability condition is \(\rho(\mathcal K)=1\).
For two-layer RBM, let \(\beta_1=\beta\) and \(\mathbf q=(u,v)\). Then
\[
M=
\begin{pmatrix}
0 & 2\beta^2\lambda_2\\
2\beta^2\lambda_1 & 0
\end{pmatrix},
\qquad
\Gamma M=
\begin{pmatrix}
0 & 2\beta^2\lambda_2\Gamma_1\\
2\beta^2\lambda_1\Gamma_2 & 0
\end{pmatrix}.
\]
Therefore by doing the algebra we find
\[
\rho(\Gamma M)
=
2\beta^2
\sqrt{
\lambda_1\lambda_2\Gamma_1\Gamma_2
},
\]
and the critical condition is therefore the following:
\begin{equation}\label{eq:appendix-RBM-splitting-critical}
4\beta^4\lambda_1\lambda_2\Gamma_1\Gamma_2=1.
\end{equation}
At zero external field, on the paramagnetic RS branch \(u=v=0\), one has \(\Gamma_1=\Gamma_2=1\), and the condition becomes the known identity \cite{BarraGenoveseGuerra2011}
\begin{equation}\label{eq:appendix-RBM-zero-field-critical}
4\beta^4\lambda_1\lambda_2=1.
\end{equation}
Equivalently, since \(\lambda_2=1-\lambda_1\), this is
\[
2\beta^2\sqrt{\lambda_1(1-\lambda_1)}=1.
\]
For a three-layer chain, the symmetric matrix \(\mathcal K\) has two non-zero adjacent entries,
\[
\mathcal K_{12}
=
2\beta_1^2
\sqrt{\lambda_1\lambda_2\Gamma_1\Gamma_2},
\qquad
\mathcal K_{23}
=
2\beta_2^2
\sqrt{\lambda_2\lambda_3\Gamma_2\Gamma_3}.
\]
Since the spectral radius of a three-node chain satisfies
\[
\rho(\mathcal K)^2=\mathcal K_{12}^2+\mathcal K_{23}^2,
\]
the splitting-instability surface is
\begin{equation}\label{eq:appendix-three-layer-splitting-critical}
4\beta_1^4\lambda_1\lambda_2\Gamma_1\Gamma_2
+
4\beta_2^4\lambda_2\lambda_3\Gamma_2\Gamma_3
=
1.
\end{equation}
This is the three-layer version of the same RS blanket-splitting criterion.

\section{Nishimori-line deformation of the CCM functional}\label{app:nishimori-deformation}

This appendix records the calculation behind the Nishimori benchmark used in Section~\ref{sec:nishimori-benchmark}.  The model studied in the main text has centered Gaussian couplings.  In the ordered-pair convention of Section~\ref{sec:nishimori-benchmark}, the Nishimori coupling between layers \(L_p\) and \(L_{p+1}\), in either orientation, is Gaussian with mean and variance tied,
\begin{equation}\label{eq:app-Nishimori-couplings}
\widetilde J_{ij}^{(p)}
=
\frac{\mu_{p,p+1}}{2N}
+
\sqrt{\frac{\mu_{p,p+1}}{2N}}\,J_{ij}^{(p)},
\qquad
J_{ij}^{(p)}\sim N(0,1).
\end{equation}
If random fields are included, the layer-\(p\) field is taken on the same Nishimori line, with mean and variance both equal to \(h^{(p)}\).  Therefore the one-body Gaussian mark is not centered.  A scalar mark of scale \(A\) has the form
\[
A+\sqrt A\,z,
\qquad z\sim N(0,1),
\]
and contributes to the functional as
\begin{equation}\label{eq:app-Nishimori-psi}
\psi(A)=\E_z\log 2\cosh(A+\sqrt A\,z).
\end{equation}
For the layer-wise CCM coordinate \(\mathbf q\), the scale seen by a spin in the architectural layer \(L_p\) is
\[
A_p(\mathbf q)=h^{(p)}+(M\mathbf q)_p.
\]
Thus the Nishimori one-body contribution is
\begin{equation}\label{eq:app-Nishimori-onebody}
\sum_{p=1}^K
\lambda_p
\psi\left(h^{(p)}+(M\mathbf q)_p\right).
\end{equation}
The correction term is modified as well.  In the centered model the RS/vector correction is
\[
\frac12(\mathbf1-\mathbf q)^\top M_1(\mathbf1-\mathbf q).
\]
The non-centered part of the Nishimori couplings produces an additional deterministic contribution on the activated adjacent-layer edge mass:
\[
-\mathbf q^\top M_1\mathbf q.
\]
Therefore the Nishimori correction is
\begin{equation}\label{eq:app-Nishimori-correction}
C_{\mathrm N}(\mathbf q)
=
\frac12(\mathbf1-\mathbf q)^\top M_1(\mathbf1-\mathbf q)
-
\mathbf q^\top M_1\mathbf q.
\end{equation}
Since \(M_1=\Lambda M_0\Lambda\), this can be written edge by edge as follows.  First,
\[
\frac12(\mathbf1-\mathbf q)^\top M_1(\mathbf1-\mathbf q)
=
\sum_{p=1}^{K-1}
\beta_p^2\lambda_p\lambda_{p+1}
(1-q^{(p)})(1-q^{(p+1)}),
\]
whereas the additional contribution
\[
\mathbf q^\top M_1\mathbf q
=
2\sum_{p=1}^{K-1}
\beta_p^2\lambda_p\lambda_{p+1}
q^{(p)}q^{(p+1)}.
\]
Hence the Nishimori correction is
\begin{equation}\label{eq:app-Nishimori-correction-edge}
C_{\mathrm N}(\mathbf q)
=
\sum_{p=1}^{K-1}
\beta_p^2\lambda_p\lambda_{p+1}
\left[
(1-q^{(p)})(1-q^{(p+1)})
-
2q^{(p)}q^{(p+1)}
\right].
\end{equation}
Combining \eqref{eq:app-Nishimori-onebody} and \eqref{eq:app-Nishimori-correction}, the Nishimori deformation of the CCM functional is
\begin{equation}\label{eq:app-CCM-Nishimori-functional}
\cP_{\mathrm N}^{\mathrm{CCM}}(\mathbf q)
=
\sum_{p=1}^K
\lambda_p
\psi\left(h^{(p)}+(M\mathbf q)_p\right)
+
\frac12(\mathbf1-\mathbf q)^\top M_1(\mathbf1-\mathbf q)
-
\mathbf q^\top M_1\mathbf q.
\end{equation}
Equivalently, by \eqref{eq:app-Nishimori-correction-edge},
\begin{align}\label{eq:app-CCM-Nishimori-functional-edge}
\cP_{\mathrm N}^{\mathrm{CCM}}(\mathbf q)
&=
\sum_{p=1}^K
\lambda_p
\psi\left(h^{(p)}+(M\mathbf q)_p\right)
\notag\\
&\quad+
\sum_{p=1}^{K-1}
\beta_p^2\lambda_p\lambda_{p+1}
\left[
(1-q^{(p)})(1-q^{(p+1)})
-
2q^{(p)}q^{(p+1)}
\right].
\end{align}

We now perform the comparison with the Nishimori variational integrand in coordinates.  In the notation of the exactly solved Nishimori DBM, the layer proportions are denoted by \(\alpha_p^{\mathrm N}\), the adjacent-layer Nishimori parameters by \(\mu_{p,p+1}\), and
\[
\Delta_{p,p+1}
=
\alpha_p^{\mathrm N}\mu_{p,p+1}\alpha_{p+1}^{\mathrm N}.
\]
The corresponding interaction matrix acts as
\[
(M^{\mathrm N}x)_p
=
\mu_{p-1,p}\alpha_{p-1}^{\mathrm N}x_{p-1}
+
\mu_{p,p+1}\alpha_{p+1}^{\mathrm N}x_{p+1},
\]
with the missing boundary terms omitted.  The finite-dimensional variational integrand on the Nishimori line is
\begin{align}\label{eq:app-known-Nishimori-integrand}
p_{\mathrm{var}}(x;\mu,h)
&=
\sum_{p=1}^K
\alpha_p^{\mathrm N}
\psi\left(h^{(p)}+(M^{\mathrm N}x)_p\right)
\notag\\
&\quad+
\sum_{p=1}^{K-1}
\frac{\Delta_{p,p+1}}2
\left[
(1-x_p)(1-x_{p+1})
-
2x_px_{p+1}
\right].
\end{align}
We now identify the Nishimori parameters with the parameters of the centered model by setting
\begin{equation}\label{eq:app-Nishimori-identification}
\alpha_p^{\mathrm N}=\lambda_p,
\qquad
\mu_{p,p+1}=2\beta_p^2,
\qquad
x_p=q^{(p)}.
\end{equation}
Then the one-body argument in equation \eqref{eq:app-known-Nishimori-integrand} becomes exactly the one-body argument in the CCM expression.  Indeed,
\[
(M^{\mathrm N}x)_p
=
2\beta_{p-1}^2\lambda_{p-1}q^{(p-1)}
+
2\beta_p^2\lambda_{p+1}q^{(p+1)}
=
(M\mathbf q)_p.
\]
Therefore we can write
\[
\sum_{p=1}^K
\alpha_p^{\mathrm N}
\psi\left(h^{(p)}+(M^{\mathrm N}x)_p\right)
=
\sum_{p=1}^K
\lambda_p
\psi\left(h^{(p)}+(M\mathbf q)_p\right).
\]
The edge coefficient also matches exactly:
\[
\frac{\Delta_{p,p+1}}2
=
\frac{\alpha_p^{\mathrm N}\mu_{p,p+1}\alpha_{p+1}^{\mathrm N}}2
=
\frac{\lambda_p(2\beta_p^2)\lambda_{p+1}}2
=
\beta_p^2\lambda_p\lambda_{p+1}.
\]
Consequently the edge part of \eqref{eq:app-known-Nishimori-integrand} becomes
\[
\sum_{p=1}^{K-1}
\beta_p^2\lambda_p\lambda_{p+1}
\left[
(1-q^{(p)})(1-q^{(p+1)})
-
2q^{(p)}q^{(p+1)}
\right],
\]
which is precisely the edge expansion \eqref{eq:app-Nishimori-correction-edge} of the CCM Nishimori correction.  Substituting these two identities into \eqref{eq:app-known-Nishimori-integrand} gives, term by term,
\begin{equation}\label{eq:app-Nishimori-final-match}
\cP_{\mathrm N}^{\mathrm{CCM}}(\mathbf q)
=
p_{\mathrm{var}}(\mathbf q;\mu,h).
\end{equation}
Thus the appendix proves the matching used in Section~\ref{sec:nishimori-benchmark}: the one-body Nishimori mark fixes the first term, the deterministic part of the non-centered couplings gives the extra contribution \(-\mathbf q^\top M_1\mathbf q\), and the identification \eqref{eq:app-Nishimori-identification} matches every adjacent-layer edge coefficient.

\section*{Acknowledgments}\label{sec:acknowledgments}
We thank Nicola Kistler (Goethe University Frankfurt) and ChatGPT (OpenAI) for important discussions and suggestions.

\section*{Declaration of Generative AI}
Research, writing, and proofreading of the present manuscript were performed with AI assistance (GPT-6 Astra) under strict human guidance and scrutiny. We independently checked all mathematical arguments and take full responsibility for the content of the publication.


\begin{thebibliography}{99}

\bibitem{SimplifiedParisiIII}
S.~Franchini,
\emph{Simplified Parisi Ansatz III: Constructive Cavity Method and Ultrametricity},
arXiv:2610.34710 (2026).
\href{https://doi.org/10.48550/arXiv.2610.34710}{doi.org/10.48550/arXiv.2610.34710}

\bibitem{RSBWithoutReplicas}
S.~Franchini,
\emph{Replica Symmetry Breaking without Replicas},
Annals of Physics \textbf{450} (2023), 169220.

\bibitem{UniversalScaling}
S.~Franchini,
\emph{Universal scaling limits for spin networks via martingale methods},
Proc. R. Soc. A \textbf{481} (2025), 20240774.
\href{https://doi.org/10.1098/rspa.2024.0774}{doi:10.1098/rspa.2024.0774}.

\bibitem{SimplifiedParisi}
S.~Franchini,
\emph{A simplified Parisi Ansatz},
Communications in Theoretical Physics \textbf{73} (5) (2021), 055601.

\bibitem{SimplifiedParisiII}
S.~Franchini,
\emph{A simplified Parisi Ansatz II: REM universality},
Chaos, Solitons \& Fractals \textbf{191} (2025), 115821.
Corrigendum: Chaos, Solitons \& Fractals \textbf{201} (2025), 117395.

\bibitem{REMUniversalityLinear}
F.~Concetti and S.~Franchini,
\emph{REM universality for linear random energy},
arXiv:2604.06122 (2026).
\href{https://doi.org/10.48550/arXiv.2604.06122}{doi.org/10.48550/arXiv.2604.06122}

\bibitem{REMUniversalityLetter}
F. Concetti and S. Franchini,
\emph{REM universality and Poisson-Dirichlet Gibbs weights for linear random energy},
arXiv:2606.07757 (2026).
\href{https://doi.org/10.48550/arXiv.2606.07757}{doi.org/10.48550/arXiv.2606.07757}

\bibitem{SherringtonKirkpatrick1975}
D.~Sherrington and S.~Kirkpatrick,
\emph{Solvable model of a spin-glass},
Phys. Rev. Lett. \textbf{35} (1975), 1792--1796.
\href{https://doi.org/10.1103/PhysRevLett.35.1792}{doi:10.1103/PhysRevLett.35.1792}.

\bibitem{Parisi1979}
G.~Parisi,
\emph{Infinite number of order parameters for spin-glasses},
Phys. Rev. Lett. \textbf{43} (1979), 1754--1756.
\href{https://doi.org/10.1103/PhysRevLett.43.1754}{doi:10.1103/PhysRevLett.43.1754}.

\bibitem{Guerra2003}
F.~Guerra,
\emph{Broken replica symmetry bounds in the mean field spin glass model},
Commun. Math. Phys. \textbf{233} (2003), 1--12.

\bibitem{Talagrand2006}
M.~Talagrand,
\emph{The Parisi formula},
Ann. of Math. \textbf{163} (2006), 221--263.

\bibitem{AizenmanSimsStarr2003}
M.~Aizenman, R.~Sims, and S.~L.~Starr,
\emph{An extended variational principle for the SK spin-glass model},
Phys. Rev. B \textbf{68} (2003), 214403.
\href{https://doi.org/10.1103/PhysRevB.68.214403}{doi:10.1103/PhysRevB.68.214403}.

\bibitem{AckleyHintonSejnowski1985}
D.~H.~Ackley, G.~E.~Hinton, and T.~J.~Sejnowski,
\emph{A learning algorithm for Boltzmann machines},
Cognitive Science \textbf{9} (1985), 147--169.

\bibitem{Smolensky1986}
P.~Smolensky,
\emph{Information processing in dynamical systems: Foundations of harmony theory},
in \emph{Parallel Distributed Processing: Explorations in the Microstructure of Cognition}, Vol.~1,
D.~E.~Rumelhart, J.~L.~McClelland, and the PDP Research Group (eds.), MIT Press, 1986, pp.~194--281.

\bibitem{SalakhutdinovHinton2009}
R.~Salakhutdinov and G.~E.~Hinton,
\emph{Deep Boltzmann machines},
Proceedings of the Twelfth International Conference on Artificial Intelligence and Statistics (AISTATS), PMLR \textbf{5} (2009), 448--455.

\bibitem{BarraGenoveseSollichTantari2017}
A.~Barra, G.~Genovese, P.~Sollich, and D.~Tantari,
\emph{Phase transitions in restricted Boltzmann machines with generic priors},
Phys. Rev. E \textbf{96} (2017), 042156.  \\
\href{https://doi.org/10.1103/PhysRevE.96.042156}{doi:10.1103/PhysRevE.96.042156}.

\bibitem{BarraContucciMingioneTantari2015}
A.~Barra, P.~Contucci, E.~Mingione, and D.~Tantari,
\emph{Multi-species mean field spin glasses. Rigorous results},
Ann. Henri Poincar\'e \textbf{16} (2015), 691--708.

\bibitem{Panchenko2015}
D.~Panchenko,
\emph{The free energy in a multi-species Sherrington--Kirkpatrick model},
Ann. Probab. \textbf{43} (2015), 3494--3513.

\bibitem{AlbericiContucciMingione2021}
D.~Alberici, P.~Contucci, and E.~Mingione,
\emph{Deep Boltzmann machines: Rigorous results at arbitrary depth},
Ann. Henri Poincar\'e \textbf{22} (2021), 2619--2642.
\href{https://doi.org/10.1007/s00023-021-01027-2}{doi:10.1007/s00023-021-01027-2}.

\bibitem{AgliariAlbaneseAlemannoFachechi2021}
E.~Agliari, L.~Albanese, F.~Alemanno, and A.~Fachechi,
\emph{Pattern recognition in Deep Boltzmann machines},
arXiv:2106.08978, 2021.

\bibitem{ChenMourrat2025}
H.-B.~Chen and J.-C.~Mourrat,
\emph{On the free energy of vector spin glasses with non-convex interactions},
Probab. Math. Phys. \textbf{6} (1) (2025), 1--80.
\href{https://doi.org/10.2140/pmp.2025.6.1}{doi:10.2140/pmp.2025.6.1}.
Preprint version: arXiv:2311.08980.

\bibitem{Chen2026}
H.-B.~Chen,
\emph{On free energy of non-convex multi-species spin glasses},
ALEA Lat. Am. J. Probab. Math. Stat. \textbf{23} (2026), 429--473.
\href{https://doi.org/10.30757/ALEA.v23-17}{doi:10.30757/ALEA.v23-17}.

\bibitem{Ruelle1987}
D.~Ruelle,
\emph{A mathematical reformulation of Derrida's REM and GREM},
Commun. Math. Phys. \textbf{108} (1987), 225--239.

\bibitem{BolthausenSznitman1998}
E.~Bolthausen and A.-S.~Sznitman,
\emph{On Ruelle's probability cascades and an abstract cavity method},
Commun. Math. Phys. \textbf{197} (1998), 247--276.
\href{https://doi.org/10.1007/s002200050450}{doi:10.1007/s002200050450}.

\bibitem{Arguin2006}
L.-P.~Arguin,
\emph{Spin glass computations and Ruelle's probability cascades},
arXiv:math-ph/0608045 (2006).

\bibitem{BarraGenoveseGuerra2011}
A.~Barra, G.~Genovese, and F.~Guerra,
\emph{Equilibrium statistical mechanics of bipartite spin systems},
J. Phys. A: Math. Theor. \textbf{44} (2011), 245002.
arXiv:1012.1261.

\bibitem{AlbericiCamilliContucciMingione2021}
D.~Alberici, F.~Camilli, P.~Contucci, and E.~Mingione,
\emph{The solution of the deep Boltzmann machine on the Nishimori line},
Comm. Math. Phys. \textbf{387} (2021), 1191--1214.
\href{https://doi.org/10.1007/s00220-021-04165-0}{doi:10.1007/s00220-021-04165-0}.

\bibitem{AuffingerChen2015}
A.~Auffinger and W.-K. Chen,
\emph{The Parisi formula has a unique minimizer},
Comm. Math. Phys. \textbf{335} (2015), 1429--1444.

\bibitem{deAlmeidaThouless1978}
J.~R.~L. de Almeida and D.~J. Thouless,
\emph{Stability of the Sherrington--Kirkpatrick solution of a spin glass model},
J. Phys. A: Math. Gen. \textbf{11} (1978), 983--990.

\bibitem{Genovese2023}
G.~Genovese,
\emph{Minimax formula for the replica symmetric free energy of deep restricted Boltzmann machines},
Ann. Appl. Probab. \textbf{33} (2023), 2324--2341.
\href{https://doi.org/10.1214/22-AAP1868}{doi:10.1214/22-AAP1868}.

\bibitem{Friston2010FEP}
K.~Friston,
\emph{The free-energy principle: a unified brain theory?},
Nature Reviews Neuroscience \textbf{11} (2010), 127--138.

\bibitem{Friston2013Life}
K.~Friston,
\emph{Life as we know it},
Journal of The Royal Society Interface \textbf{10} (2013), 20130475.

\bibitem{Kirchhoff2018MarkovBlankets}
M.~D. Kirchhoff, T.~Parr, E.~Palacios, K.~Friston, and J.~Kiverstein,
\emph{The Markov blankets of life: autonomy, active inference and the free energy principle},
Journal of The Royal Society Interface \textbf{15} (2018), 20170792.

\bibitem{FranchiniBardella2026}
S.~Franchini, G.~Bardella,
\emph{Lattice Field Theory for a network of real neurons},
42nd International Symposium on Lattice Field Theory (LATTICE2025), arXiv:2604.05251 (2026).

\bibitem{BardellaFranchini2024}
G.~Bardella, S.~Franchini, L.~Pan, R.~Balzan, E.~Brunamonti, P.~Pani, S.~Ferraina,
\emph{Neural activity in quarks language: Lattice Field Theory for a network of real neurons},
Entropy \textbf{26} (2024), 495.

\bibitem{BardellaFranchini2025}
G.~Bardella, S.~Franchini, P.~Pani, S.~Ferraina,
\emph{Lattice physics approaches for neural networks},
iScience \textbf{27} (2024), 111390.

\bibitem{HartnettParkerGeist2018}
G.~S.~Hartnett, E.~Parker, and E.~Geist,
\emph{Replica symmetry breaking in bipartite spin glasses and neural networks},
Phys. Rev. E \textbf{98} (2018), 022116.
\href{https://doi.org/10.1103/PhysRevE.98.022116}{doi:10.1103/PhysRevE.98.022116}.
Cited preprint version:
\href{https://arxiv.org/abs/1803.06442v4}{arXiv:1803.06442v4}.

\bibitem{ChenIssaMourrat2026}
H.-B.~Chen, V.~Issa, and J.-C.~Mourrat,
\emph{Free energy of non-convex multi-species spin glasses with centered Ising spins},
\href{https://arxiv.org/abs/2606.16636v2}{arXiv:2606.16636v2} (2026).

\end{thebibliography}
\end{document}